\documentclass[twocolumn]{aastex62}

\usepackage{rotating}
\usepackage{natbib}

\begin{document}

\title{Predictions of DKIST/DL-NIRSP observations for an off-limb kink-unstable coronal loop}



\author{B. Snow}
\affiliation{School of Mathematics and Statistics, University of Sheffield, S3 7RH, UK}
\affiliation{Northumbria University, Newcastle upon Tyne, NE1 8ST, UK}
\affiliation{\added{University of Exeter, Exeter, EX4 4QF, UK}}

\author{G. J. J. Botha}
\affiliation{Northumbria University, Newcastle upon Tyne, NE1 8ST, UK}

\author{E. Scullion}
\affiliation{Northumbria University, Newcastle upon Tyne, NE1 8ST, UK}

\author{J. A. McLaughlin}
\affiliation{Northumbria University, Newcastle upon Tyne, NE1 8ST, UK}

\author{P. R. Young}
\affiliation{NASA Goddard Space Flight Center, Greenbelt, MD 20771, USA}
\affiliation{George Mason University, Fairfax, VA 22030, USA}
\affiliation{Northumbria University, Newcastle upon Tyne, NE1 8ST, UK}

\author{S. A. Jaeggli}
\affiliation{National Solar Observatory, 8 Kiopa`a St. Suite 201, Pukalani, HI 96768 USA}

\begin{abstract}
Synthetic intensity maps are generated from a 3D kink-unstable flux rope simulation using several DKIST/DL-NIRSP spectral lines to make a prediction of the observational signatures of energy transport and release. The reconstructed large field-of-view intensity mosaics and single tile sit-and-stare high-cadence image sequences show detailed, fine-scale structure and exhibit signatures of wave propagation, redistribution of heat, flows and fine-scale bursts. These fine-scale bursts are present in the synthetic Doppler velocity maps and can be interpreted as evidence for small-scale magnetic reconnection at the loop boundary. 
The spectral lines reveal the different thermodynamic structures of the loop, with the hotter lines showing the loop interior and braiding, and the cooler lines showing the radial edges of the loop.
The synthetic observations of DL-NIRSP are found to preserve the radial expansion and hence the loop radius can be measured accurately.
The electron number density can be estimated using the intensity ratio of the Fe \textsc{xiii} lines at 10747 and 10798 \AA. The estimated density from this ratio is correct to within $\pm 10 \%$ during the later phases of the evolution, however \added{it} is less accurate initially when \added{line-of-sight} density inhomogeneities contribute to the Fe \textsc{xiii} intensity, resulting in an overprediction of the density by $\approx 30 \%$. 
The identified signatures are all above a conservative estimate for instrument noise and therefore will be detectable. \added{In summary,} we have used forward modelling to demonstrate that the coronal off-limb mode of DKIST/DL-NIRSP will be able to detect multiple independent signatures of a kink-unstable loop and observe small-scale transient features including loop braiding/twisting and small-scale reconnection events occurring \added{at} the radial edge of the loop.  

\vspace{1cm}
\end{abstract}

\section{Introduction} \label{sec:intro}

\deleted{The temperature in the solar corona is thought to be maintained primarily through a cascade of reconnection events referred to as nanoflares \citep{Parker1988}.}
\added{One mechanism capable of producing the high temperatures observed in the solar corona is nanoflares, where a cascade of reconnection events releases magnetic energy, leading to heating \citep{Parker1988}.} 
Such heating events are assumed to be the result of reconnection occurring in tangled or twisted magnetic fields \citep[e.g.][]{Parker1988,Hood2009,Bareford2015}. 
\added{The origin of twisted coronal fields is thought to be due to photospheric convective/shearing motions which propagate into the corona due to the relatively long coronal diffusion times, resulting in magnetic reconnection \citep[][]{Parker1988,Pariat2015,Kumar2017}. Alternatively, twisted flux can emerge from below the photosphere \citep[][]{Ishii1998,Cheung2014,Takasao2015}. 
} 
These twisted magnetic structures are susceptible to instabilities, such as the kink instability, whereby magnetic energy is released creating pressure gradients, flows and local temperature increases. However, convincing evidence of nanoflares and small-scale reconnection events remains elusive in observations from current ground-based and space-borne instruments \citep{Parnell2012}. In this paper, we consider the observational signatures of energy transport from a simulation of a kink-unstable flux rope via forward modelling using the off-limb coronal mode of the forthcoming DKIST/DL-NIRSP instrument. 

The non-eruptive kink instability has been used to trigger reconnection in numerical simulations of cylindrical coronal loops \citep{Browning2008,Botha2011kink,Gordovskyy2016,Pinto2016}. The general process is as follows: [1] a magnetic field is specified that is unstable to the ideal kink-instability, [2] the simulation is allowed to evolve ideally resulting in an increase in the twist of the magnetic field, [3] at a critical value, anomalous resistivity is activated and the reconnection occurs at current sheets, producing local temperature increases, flows and reducing the overall twist present in the simulation. The majority of the heating is a result of shock heating as opposed to Ohmic heating \citep{Bareford2015,Bareford2016}. Localised heating is conducted away from reconnection sites parallel to magnetic field lines through thermal conduction, reducing the peak temperatures obtained in the loop \citep{Botha2011kink}.

To accurately compare simulation results to observations, and to predict signatures of events, it is necessary to apply forward modelling. This is a process whereby the numerical data is converted into observables such as line-of-sight \added{(LOS) integrated} intensity and Doppler velocities. For coronal phenomena, the plasma is usually assumed to be optically thin and, as such, only in emission. This results in the observable intensity and Doppler velocity being a function of temperature, density and velocity from the numerical simulations. This methodology has been applied to synthesis of the observational signatures of various solar phenomena \citep[e.g.][]{Verwichte2009,Peter2012,Demoortel2015,Snow2015,Mandal2016,Yuan2016,Snow2017}. 
Synthetic observables using several SDO/AIA and TRACE broadband EUV filters reveal the general twisted structure of a kink-unstable loop \citep{Botha2012,Srivastava2013}, and there are several signatures of energy release and energy transport when observing through various Hinode/EIS lines \citep{Snow2017}.
However, the forward modelling does not show clear signatures of nanoflares or sites of small-scale reconnection, even when they are present in the numerical simulation. \added{This is because the} spatial and temporal degradation for these instruments result in small-scale, transient features becoming averaged out and are not visible in the synthetic observables. 

In this paper, the forward modelling of a 3D numerical simulation of a non-eruptive kink-unstable flux rope is performed using spectral lines of the coronal off-limb mode of DKIST/DL-NIRSP. Intensity and Doppler maps are generated and the results are compared to the simulation.  Both the mosaic and sit-and-stare observing modes (see Section \ref{results} below) of the low-resolution off-limb coronal channel of DKIST/DL-NIRSP are investigated. The signal-to-noise ratio \added{(SNR)} and photon rates are estimated for the different spectral lines to ensure the observability of the results. This allows us to make a prediction of the observational signatures of the off-limb coronal mode of DKIST/DL-NIRSP. 

\section{Instrument information} 

\begin{table*}
\caption{Spatial Properties of DL-NIRSP}
\centering
\begin{tabular}{rccc}
Property & High Resolution & Mid Resolution & Wide Field \\
\hline
Focal Ratio & f/62 & f/24 & f/8 \\
Plate Scale [$''$/mm] & 0.832 & 2.15 & 6.45 \\
BiFOIS Field of View Width $\times$ Height [$''$] & 2.40 $\times$ 1.80 & 6.19 $\times$ 4.64 & 18.6 $\times$ 27.8 \\
Maximum Field of View [$''$] & 120 & 120 & 120 \\
Instrument Spatial Sampling[$''$] & 0.03 & 0.077 & 0.464
\end{tabular}
\label{tab:spatialproperties}
\end{table*}


The Diffraction Limited Near Infrared Spectropolarimeter (DL-NIRSP) is an imaging spectropolarimeter currently under development by the University of Hawai`i's Institute for Astronomy for the Daniel K. Inouye Solar Telescope (DKIST), which will see first light early in 2020.  DL-NIRSP will have the unique ability to obtain high cadence spectropolarimetric measurements of small fields of view through the use of a bi-dimensional fibre optic image slicer (BiFOIS).  DL-NIRSP will support disk, limb, and off-disk coronal observations of up to three simultaneous wavelength regions in the visible and near infrared.  The selected diagnostic lines are targeted at making multi-height magnetic field measurements and are planned to include \ion{Fe}{11} 7892, \ion{Ca}{2} 8542, \ion{Fe}{13} 10747, \ion{He}{1} 10830, \ion{Si}{10} 14300, and \ion{Fe}{1} 15650 \AA.

The DL-NIRSP feed optics provide three different resolution modes:  high resolution f/62, mid resolution f/24, and wide field f/8.  The first two modes are suitable for disk and limb observations, while the wide-field mode is primarily for coronal observations. \added{Details of the spatial properties of DL-NIRSP are given in Table \ref{tab:spatialproperties}.} \deleted{ Two different format BiFOIS will be used for the on-disk and off-disk modes.  Each of the BiFOIS units will be constructed of two adjacent stacks of fibre ribbons, where each ribbon is composed of many fibres.  The on-disk BiFOIS device will cover an area of $2.5''\times1.8''$ at f/62 and $6.5''\times4.6''$ at f/24 mode, while the off-disk BiFOIS device, with $2\times$ larger fibre cores, will cover $18.6''\times27.8''$ at f/8.  A precision steerable mirror in the feed optics makes it possible to mosaic this small field up to the full field of 2.8 arcmin diameter. The large rectangular fibre cores ensure that linear polarization states modulated by the rotating waveplate immediately upstream of BiFOIS are maintained.  The close packed array at the input of BiFOIS is reformatted into five evenly spaced slits at the entrance to the spectrograph.}

\deleted{The spectrograph design is an off-axis Littrow configuration similar to the Facility Infrared Spectropolarimeter (FIRS) on the Dunn Solar Telescope \citep{Jaeggli2010}.  Collimated light is fed to a grating with 23.2 lines/mm and a blaze angle of $63^\circ$.  Following the grating the dispersed beam is distributed between three camera arms, one for the visible and two for the infrared, where the beam is analysed by a Wollaston prism which splits the light into two linearly polarized beams.  Interchangeable bandpass isolation filters specific to each wavelength prevent light from adjacent slits and other orders from overlapping at the detector.  A 4k Andor CCD detector will be used for the visible arm, and Hawaii H2RG 2k HgCdTe arrays will be used for each of the infrared arms.  The visible and IR cameras will be capable of a frame rate of at least 30 Hz.}

\deleted{DL-NIRSP is situated at the end of the facility light distribution optics and as such it will receive a light feed corrected by the high order adaptive optics system and will be able to observe simultaneously with the other instruments with the exception of Cryo-NIRSP.}

\section{Computational model and loop evolution}

\begin{table}
\centering
\caption{Initialisation in the numerical simulation}
\begin{tabular}{l | l}
Temperature & 0.125 MK  \\
Density     & $1.67 \times 10^{-12}$ kg m$^{-3}$ \\
Electron number density & $10^{9}$ cm$^{-3}$ \\
\hline
Loop length & 80 Mm \\
Loop radius & 4 Mm \\
\hline
Magnetic field & \\
\hspace{5mm}Inside loop & 20 G maximum \\
\hspace{5mm}Outside loop & 15 G uniform \\
\hspace{5mm}Twist & 0 at axis and loop edge \\
                  & $11.5\pi$ at radius 1 Mm \\
\hline
Numerical & \\
Grid: $x$ axis & [ -8 Mm, 8 Mm ] \\
\phantom{Grid:} $y$ axis & [ -8 Mm, 8 Mm ] \\
\phantom{Grid:} $z$ axis & [ 0, 80 Mm ] \\
Cell: $\delta x = \delta y$ & 0.125 Mm \\
\phantom{Cell:} $\delta z$  & 0.3125 Mm \\
Boundaries: $x$ and $y$   & reflective \\
\phantom{Boundaries:} $z$ & line-tied 
\end{tabular}
\label{simspec}
\end{table}

The 3D numerical simulation was performed by \cite{Botha2011kink} using Lare3d \citep{Arber2001}, solving the resistive magnetohydrodynamic (MHD) equations with Spitzer-H\"arm thermal conductivity acting parallel to magnetic field lines. \added{The initial conditions of the simulation are based on the observation of \cite{Srivastava2010}.} The loop is initialised as a straight cylinder with aspect ratio 10 (Table \ref{simspec}) and with a force-free magnetic field that is unstable to the ideal MHD kink instability. The initial magnetic field twist is above the critical numerical stability limit of $4.8\pi$ \citep{Mikic1990}. The initial temperature and density are uniform with values given in Table \ref{simspec} and the initial electron number density $n_e$ is obtained assuming quasi-neutrality. The numerical domain is a Cartesian box with the loop axis along the $z$ direction. The $x$ and $y$ boundaries are far enough from the loop's radial edge not to influence the results. The $z$ boundary condition at the \added{loop} footpoints has no velocity component across it and has constant density and temperature values fixed at their initial values. A temperature gradient is allowed so that heat flows across the boundary. Full details of the numerical model can be found in \cite{Botha2011kink}.

The time evolution of the kink instability can be separated into two phases: a linear phase during which current sheets form and grow, and a non-linear phase (starting at time $t=261$ seconds) where reconnection occurs in current sheets. This reconnection releases magnetic energy causing localised heating and the straightening of magnetic field lines. The local heating results in pressure gradients that act to drive flows along magnetic field lines, while thermal conduction acts to equalise the temperature inside the loop. A full description of the physics during the time evolution is given in \cite{Botha2011kink}.

\section{Forward modelling and line synthesis}

\begin{figure}
\centering
\includegraphics[scale=0.5,clip=true, trim=3cm 7cm 3cm 8cm]{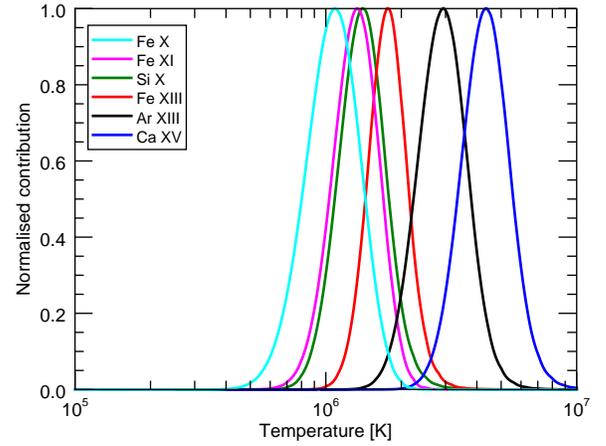}
\caption{Contribution functions for the synthesised spectral lines.}
%
\label{figresp}
\end{figure}

The simulation of the kink-unstable coronal loop has been used previously to generate observational images for TRACE 171 \AA~and SDO/AIA \citep{Botha2012,Srivastava2013} as well as for Hinode/EIS \citep{Snow2017}, in order to investigate the signatures of energy release and transport during the evolution of the non-linear phase of the kink instability. In this paper, we follow a similar approach for DKIST/DL-NIRSP, where the synthetic intensity is calculated according to
\begin{equation}
I = \int n_e ^2 C(T) \: dl,
\label{eqnintens}
\end{equation}
where $n_e$ is the electron number density, $C(T)$ is the contribution function, $T$ is temperature and $dl$ is the unit length along the line-of-sight (LOS). This formula uses the optically thin assumption, whereby it is assumed that the plasma is entirely in emission. This assumption is valid for coronal temperatures such as those studied here. The electron density $n_e$ is provided by the numerical simulation using the assumption of quasi-neutrality, and the contribution functions $C(T)$ are synthesised using CHIANTI v8. The emission is then integrated along the LOS to produce 2D intensity maps for the different spectral lines. The signal is spatially integrated to the instrument pixel size and temporally integrated to the instrument exposure time.

The contribution functions from several spectral lines that can be observed using DKIST/DL-NIRSP have been synthesised using CHIANTI v8 \citep{dere1997chianti,DelZanna2015}. Spectral information about these lines is given in Table \ref{tabspec}. \added{The contribution functions were synthesised using the initialisation density for the simulation. The effect of density on these contribution functions was found to be negligible for the intensity and Doppler maps. For the density diagnostic investigated in Section \ref{sec:dens}, the contribution functions were resynthesised in each simulation cell using the local density and temperature values.} Note that photoexcitation from the photospheric radiation field is omitted in the line synthesis since it was found to be negligible for the considered spectral lines. In this paper, we investigate the observational signatures using the Fe \textsc{xi} (7892 \AA), Fe \textsc{xiii} (10747 \AA) and Ca \textsc{xv} (5695 \AA) spectral lines, covering a large thermal range of the loop. The contribution functions for these lines are shown in Figure \ref{figresp}. Note that the plot also contains the contribution function for several lines that were initially considered for DKIST/DL-NIRSP but are not included in the current specification.

DKIST/DL-NIRSP will observe the three coronal lines of Fe \textsc{xi}, Fe \textsc{xiii} and Si \textsc{x} at first light.
The results from the Si \textsc{x} line are very similar to those from the Fe \textsc{xi} line as the two lines are formed at the same temperature, and so they are not shown here. The Fe \textsc{xi} line also has a higher predicted photon rate than the Si \textsc{x} line, see Section \ref{sec:snr}.
Instead we present the observables from the Ca \textsc{xv} line, a previously considered line that was omitted in the re-specification. This activates at a higher temperature than the Fe \textsc{xi} and Fe \textsc{xiii} lines and therefore shows different thermal structures. We believe there is significant benefit to including this line in future re-specifications of DKIST/DL-NIRSP. Note that the Ca \textsc{xv} line will be observed using the DKIST/VISP instrument.

\begin{table}
\centering
\caption{Spectral information for the different spectral lines. Fe \textsc{xi}, Fe \textsc{xiii} and Ca \textsc{xv} are used in this paper. \added{The wavelengths are given as their rest values in air.}}
\begin{tabular}{p{1.6cm} | p{2.4cm} | p{2.1cm}}
Spectral line    & Air wavelength (\AA) & Temperature peak (log(T)) \\
               \hline
Fe {\sc{x}}     & 6375          & 6.05 \\
Fe {\sc{xi}}     & 7892          & 6.15 \\
Si {\sc{x}}      & 14300         & 6.15 \\
Fe {\sc{xiii}}   & 10747         & 6.25 \\
Fe {\sc{xiii}}   & 10798         & 6.25 \\
Ar {\sc{xiii}}   & 10140         & 6.45 \\
Ca {\sc{xv}}     & 5695          & 6.65 
\end{tabular}
\label{tabspec}
\end{table}

The wide-field mode of DKIST/DL-NIRSP is used for this paper\added{, see Table \ref{tab:spatialproperties}}. Each mosaic tile focuses on off-limb observations of coronal lines and has a sampling size of $ 0.464''$ and a field of view (FOV) of $18.6'' \times 27.8''$. The medium and high resolution channels are targeted at bright chromospheric and photospheric regions on the solar disk and would require longer exposure times in the corona, they are therefore less useful for this study, where we consider the dynamic signatures of a coronal flux rope. 

In this study, we use conservative estimates for the spatial resolution ($ 0.464''$) and exposure time (22.9 s).
Under good instrumental and atmospheric scattered light conditions, far higher spatial and temporal resolution will be possible. This paper presents results using fairly pessimistic values of the scattered light in order to assess a low-quality observation of a kink-unstable coronal loop using DKIST/DL-NIRSP. In this way, we identify a minimum threshold that \added{will} be detectable. Various observational signatures of this event are described and the observability is tested by consideration of the signal-to-noise ratio and photon rates.

\section{Results}
\label{results}

We consider two types of image sequence: mosaic and sit-and-stare. For the mosaic configuration, three tiles (of $18.6'' \times 27.8''$) are scanned sequentially (with an exposure time of 22.9 s) from left to right in the image plane, resulting in a larger effective FOV (of $55.7'' \times 27.8''$) and this mosaic image sequence is repeated resulting in an effective mosaic cadence of $69$ s. In the sit-and-stare mode, a single mosaic tile is observed repeatedly without moving in the image plane resulting in a higher cadence image sequence but with a reduced effective FOV ($18.6'' \times 27.8''$). 

\subsection{Mosaic configuration}

\begin{figure*}
\begin{tabular}{p{5.5cm} p{2.7cm} p{2.6cm} p{5cm} p{1cm}}
 & left & centre & right & 
\end{tabular} 
\vspace{-0.6cm}
{\center
\includegraphics[scale=0.08,clip=true,trim=15cm 0cm 15cm 0cm]{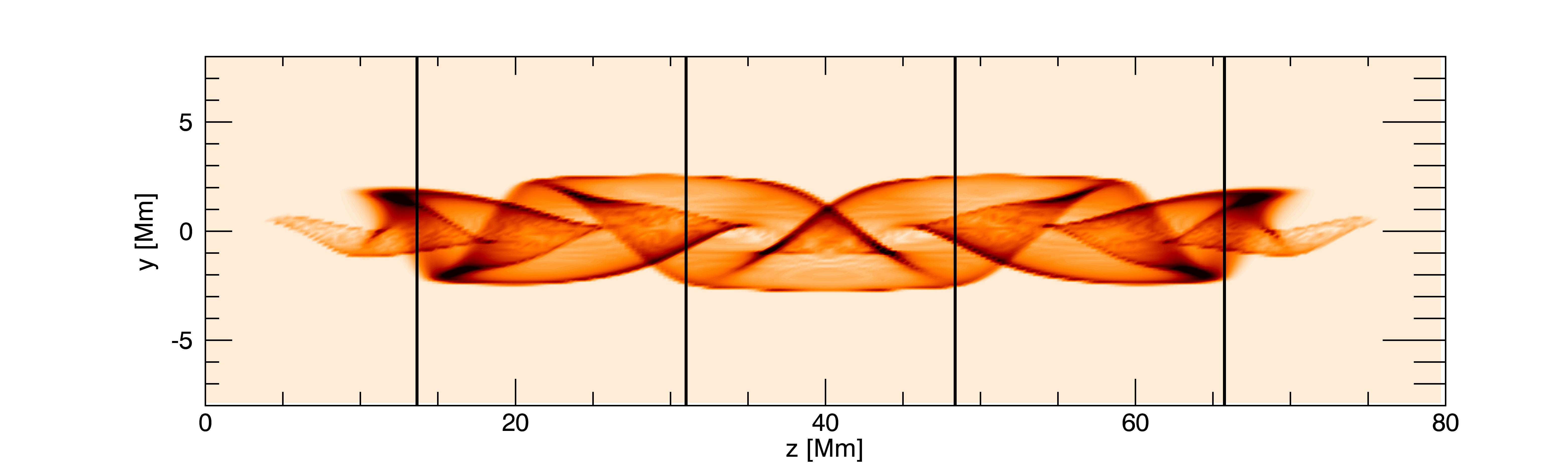}
\caption{Contour of the intensity of the loop at time $t=261$ seconds using the Fe \textsc{xiii} (10747 \AA) spectral line. The four black lines mark the edge of the three mosaic tiles investigated in this section.}}
\label{figtiles}
\end{figure*}

We consider three tiles, each with a FOV of $18.6 \arcsec \times 27.8 \arcsec$, referred to as the coronal mode of DL-NIRSP. The three-tile scan is centred on the middle of the loop, as shown in Figure \ref{figtiles}. We will refer to these three tiles individually as the left, centre and right tiles. These tiles form a simple mosaic, with a scan sequence order of left to centre to right. This sequence is suitable for this simulation as it captures the majority of \added{the} length of the loop. The FOV does not capture the loop footpoints but instead focuses on the central part of the loop which has much more dynamic behaviour. Note, the simulation snapshot presented in Figure \ref{figtiles} is taken at the full numerical spatial and temporal resolution. In contrast, the tiles will be exposed with instrumental exposure time resulting in blurring of the structures present at full numerical resolution. For example, the exposure time frames (in seconds) for the tiles for the first mosaic are $261 \leq t < 284$ for the left tile, $284 \leq t < 307$ for the centre tile, and $307 \leq t < 330$ for the right tile, then the imaging sequence restarts at the left tile position. This mosaic sequence is repeated until the end of the simulation time. Thus, this three-tile mosaic configuration has an effective cadence of 69~s, i.e.\ (330 - 261)~s. 

\subsubsection{Intensity mosaic}

The intensity along a \added{LOS} is calculated according to Equation (\ref{eqnintens}). This intensity is then averaged over the DKIST/DL-NIRSP pixel size and integrated over time to match the instrumental exposure time (22.9 seconds).

\begin{figure*}
\vspace{0cm}
\begin{tabular}{c c}
& \hspace{-2cm} x \hspace{8cm} y \\
\begin{turn}{90} \hspace{0.8cm} $t=261$\end{turn}& \includegraphics[scale=0.07]{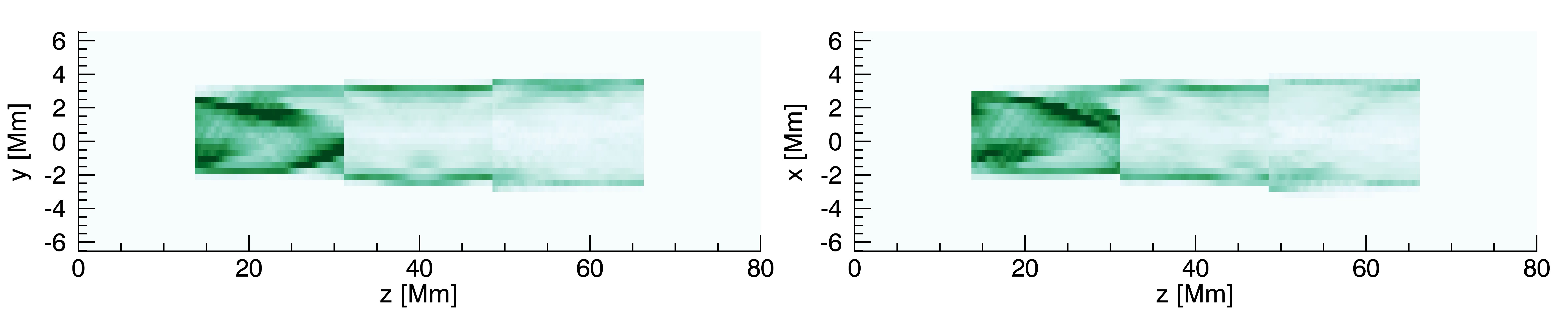} \\
\begin{turn}{90} \hspace{0.8cm} $t=330$\end{turn}& \includegraphics[scale=0.07]{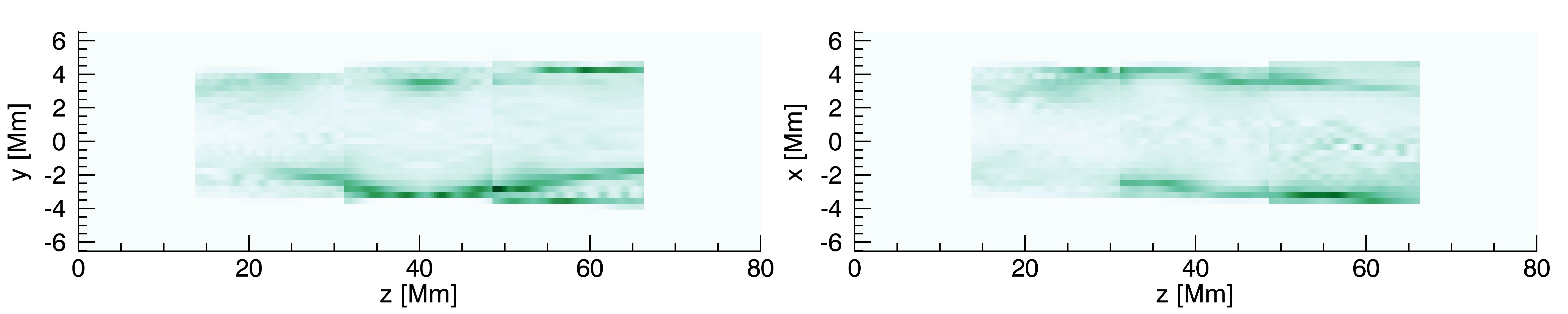} \\
\begin{turn}{90} \hspace{0.8cm} $t=400$\end{turn}& \includegraphics[scale=0.07]{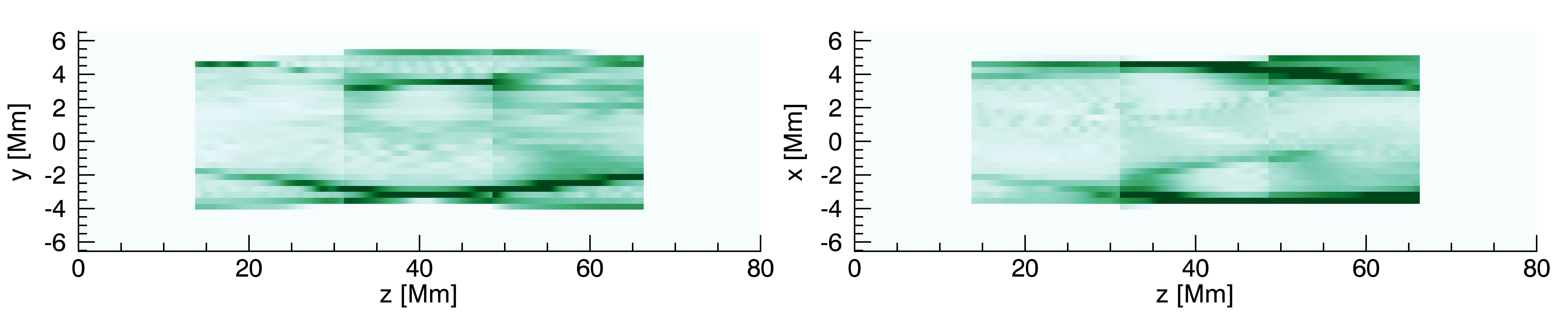} \\
\begin{turn}{90} \hspace{0.8cm} $t=469$\end{turn}& \includegraphics[scale=0.07]{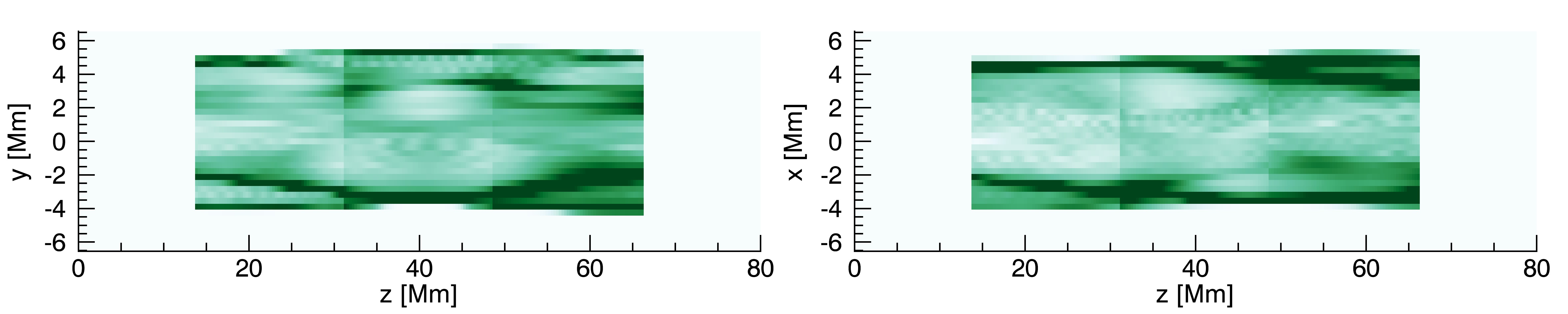} \\
\begin{turn}{90} \hspace{0.8cm} $t=539$\end{turn}& \includegraphics[scale=0.07]{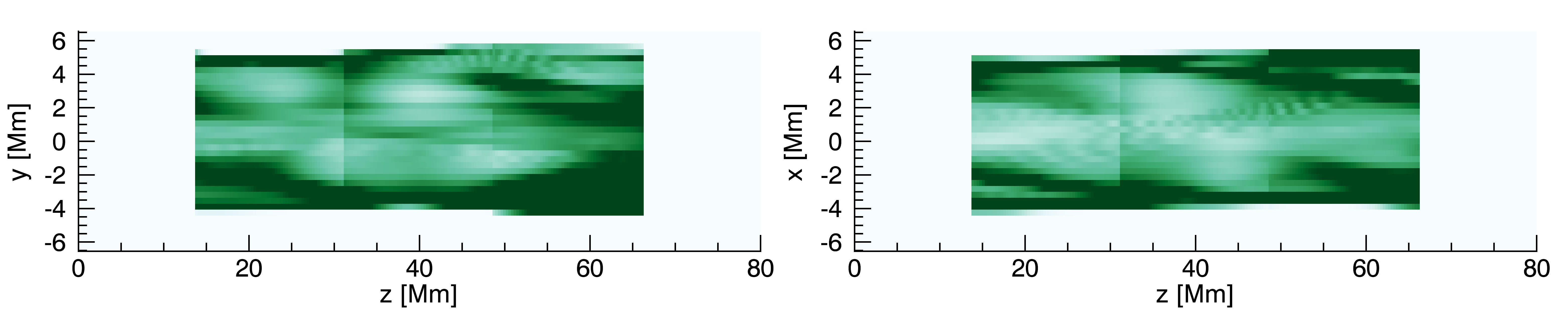} \\
\begin{turn}{90} \hspace{0.8cm} $t=609$\end{turn}& \includegraphics[scale=0.07]{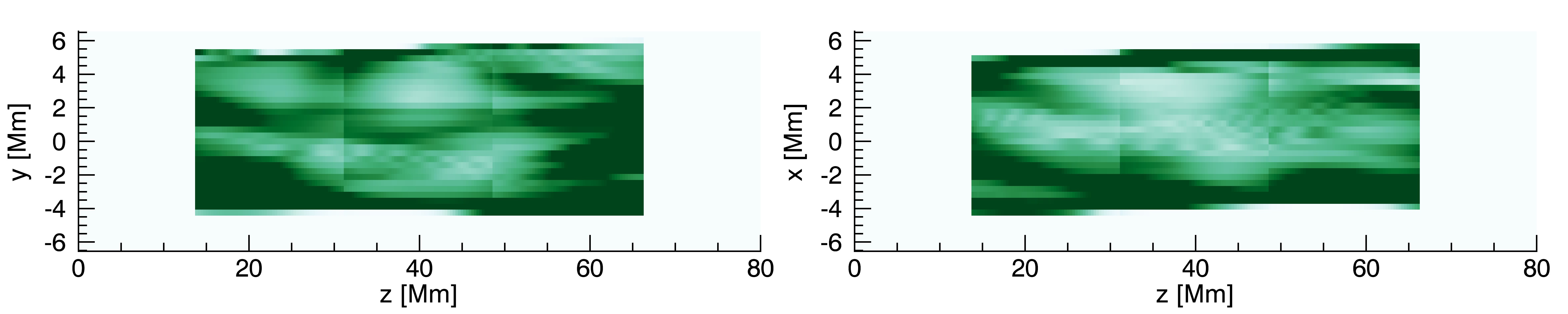} \\
\begin{turn}{90} \hspace{0.8cm} $t=679$\end{turn}& \includegraphics[scale=0.07]{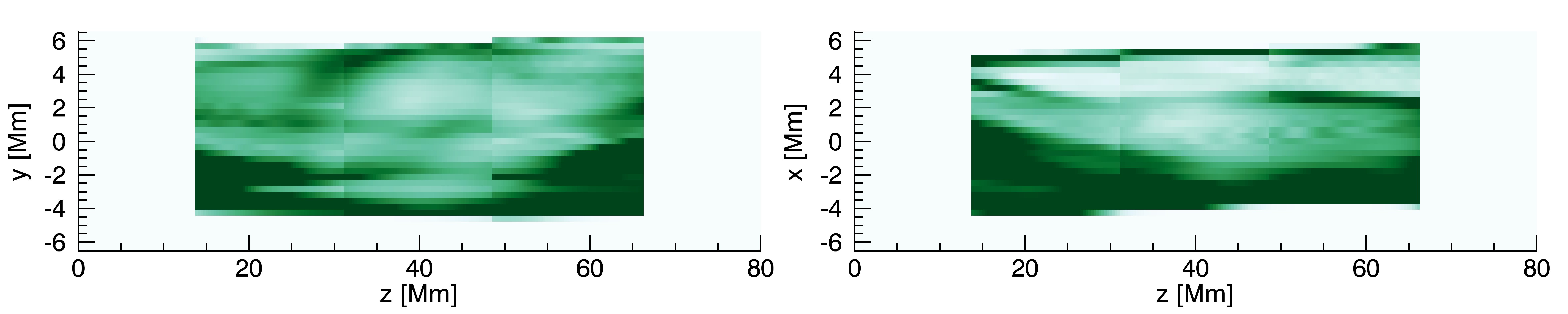}
\end{tabular}
\caption{Intensities Fe \textsc{xi} using a moving mosaic \added{imaging sequence}, integrated along the $x$ (left) and $y$ (right) directions. Dark colour indicates high intensity.}
\label{figmosfe11}
\end{figure*}

\begin{figure*}
\vspace{0cm}
\begin{tabular}{c c}
& \hspace{-2cm} x \hspace{8cm} y \\
\begin{turn}{90} \hspace{0.8cm} $t=261$\end{turn}& \includegraphics[scale=0.07]{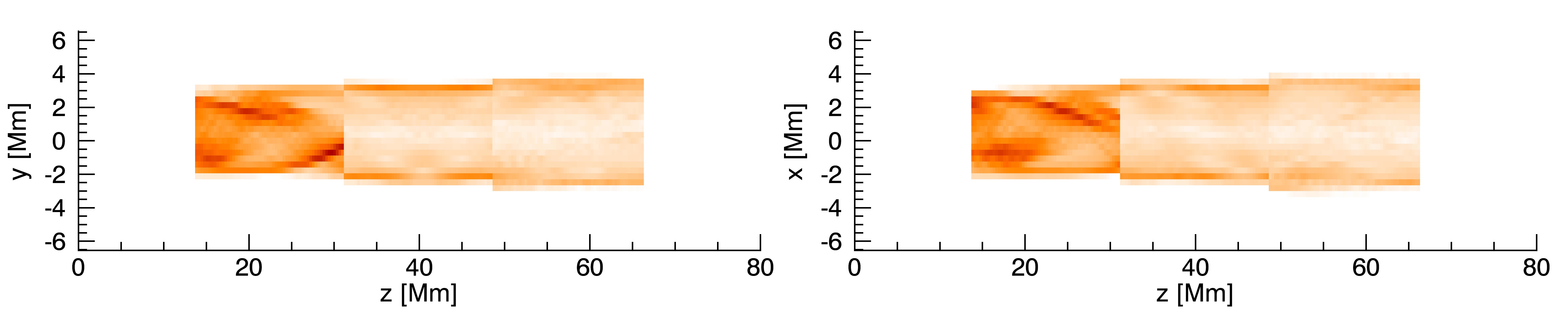} \\
\begin{turn}{90} \hspace{0.8cm} $t=330$\end{turn}& \includegraphics[scale=0.07]{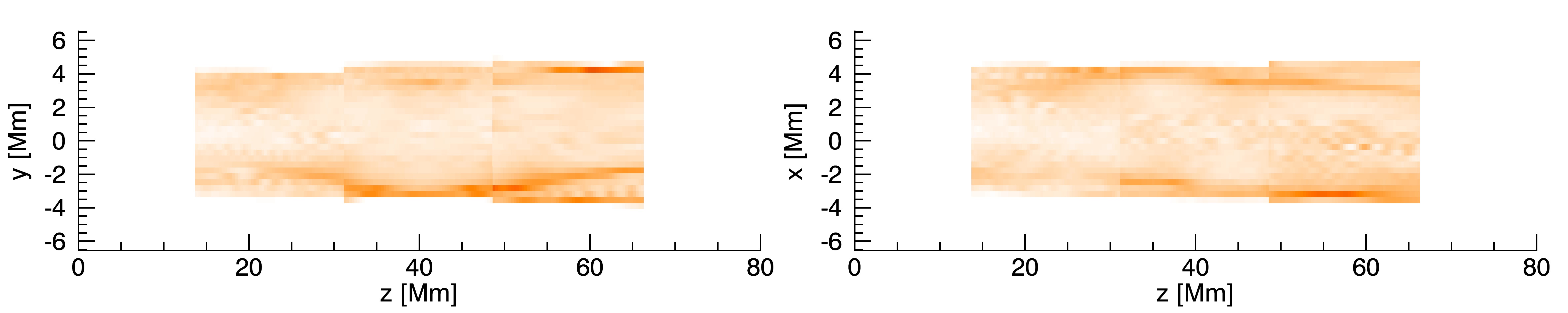} \\
\begin{turn}{90} \hspace{0.8cm} $t=400$\end{turn}& \includegraphics[scale=0.07]{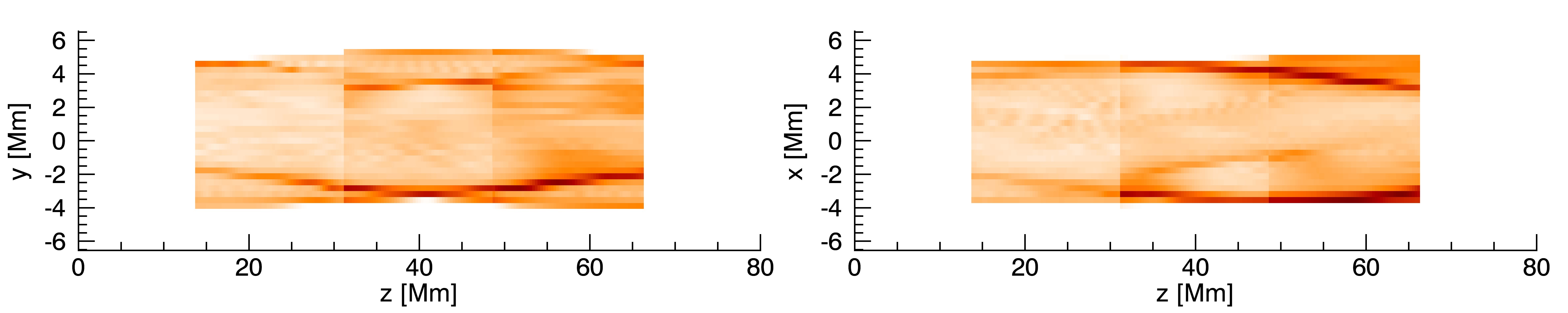} \\
\begin{turn}{90} \hspace{0.8cm} $t=469$\end{turn}& \includegraphics[scale=0.07]{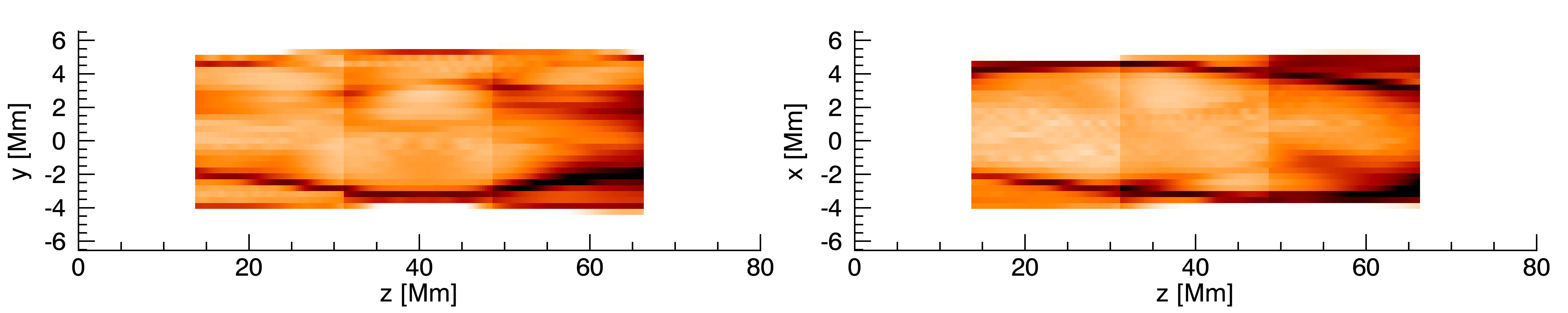} \\
\begin{turn}{90} \hspace{0.8cm} $t=539$\end{turn}& \includegraphics[scale=0.07]{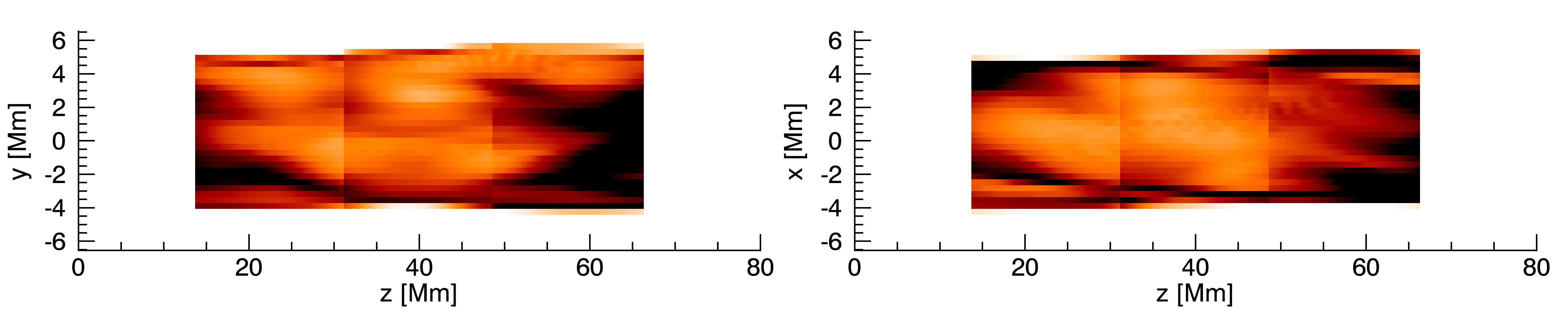} \\
\begin{turn}{90} \hspace{0.8cm} $t=609$\end{turn}& \includegraphics[scale=0.07]{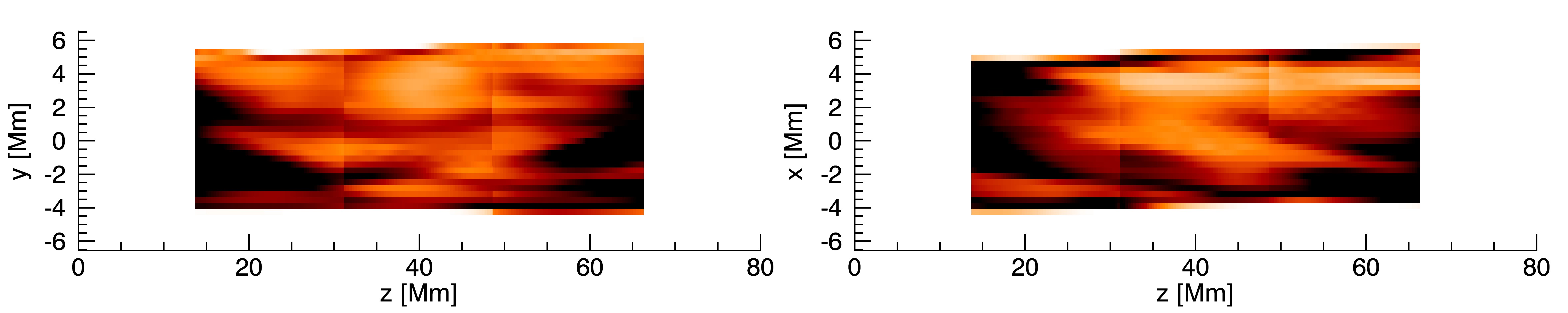} \\
\begin{turn}{90} \hspace{0.8cm} $t=679$\end{turn}& \includegraphics[scale=0.07]{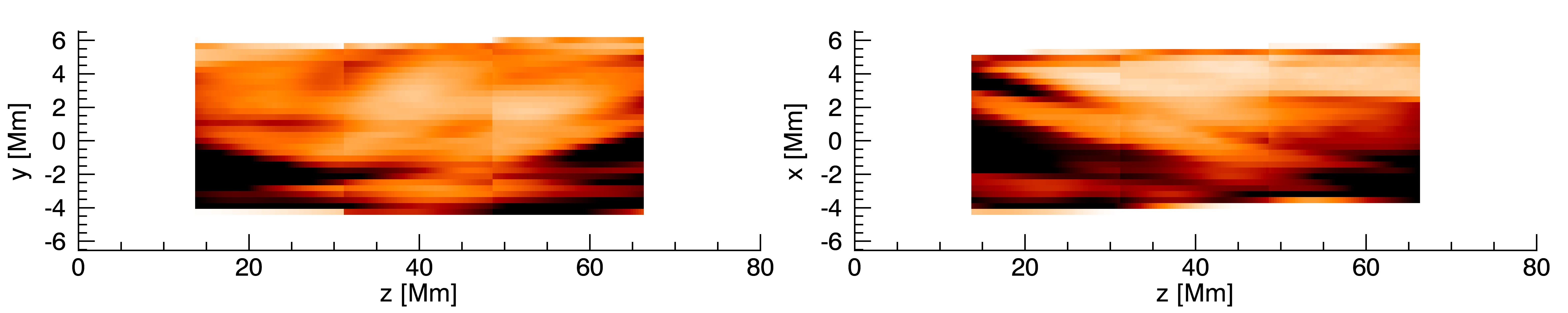}
\end{tabular}
\caption{Intensities Fe \textsc{xiii} using a moving mosaic \added{imaging sequence}, integrated along the $x$ (left) and $y$ (right) directions. Dark colour indicates high intensity.}
\label{figmosfe13}
\end{figure*}

\begin{figure*}
\vspace{0cm}
\begin{tabular}{c c}
& \hspace{-2cm} x \hspace{8cm} y \\
\begin{turn}{90} \hspace{0.8cm} $t=261$\end{turn}& \includegraphics[scale=0.07]{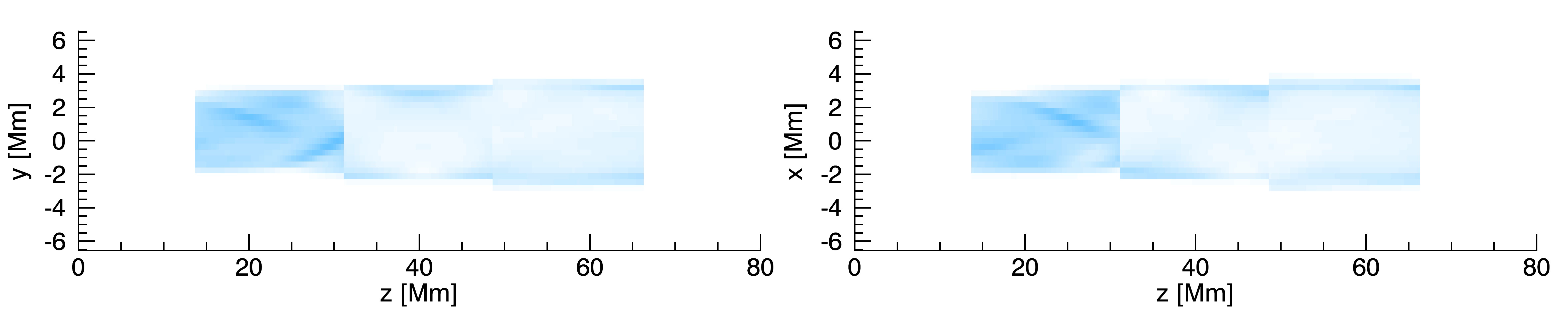} \\
\begin{turn}{90} \hspace{0.8cm} $t=330$\end{turn}& \includegraphics[scale=0.07]{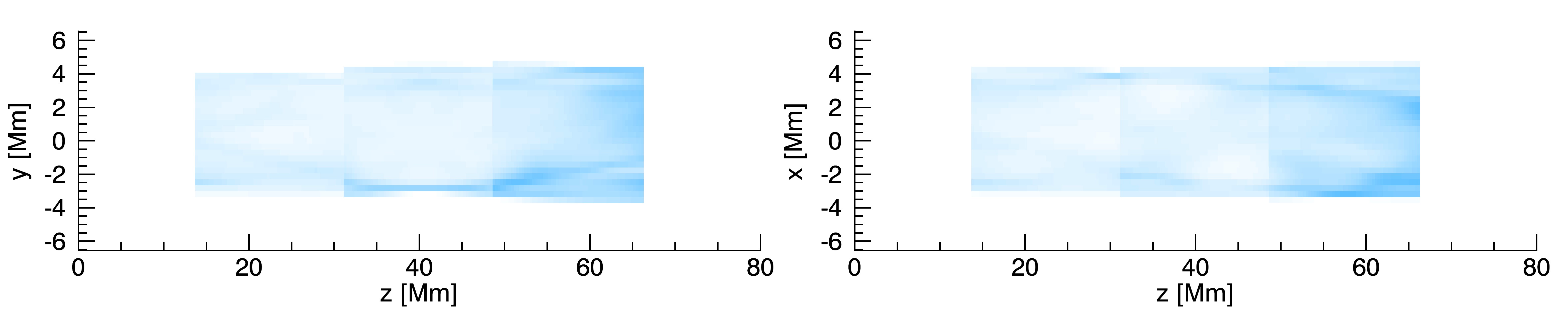} \\
\begin{turn}{90} \hspace{0.8cm} $t=400$\end{turn}& \includegraphics[scale=0.07]{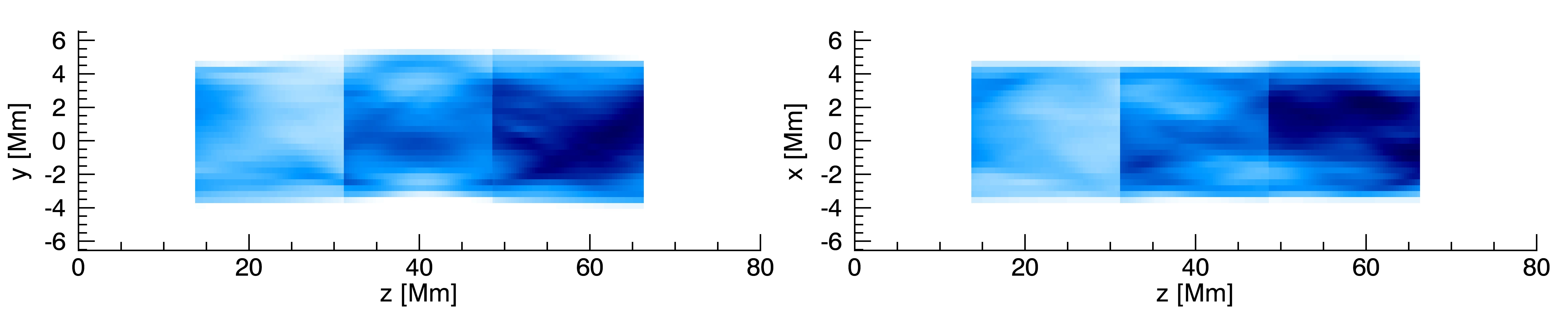} \\
\begin{turn}{90} \hspace{0.8cm} $t=469$\end{turn}& \includegraphics[scale=0.07]{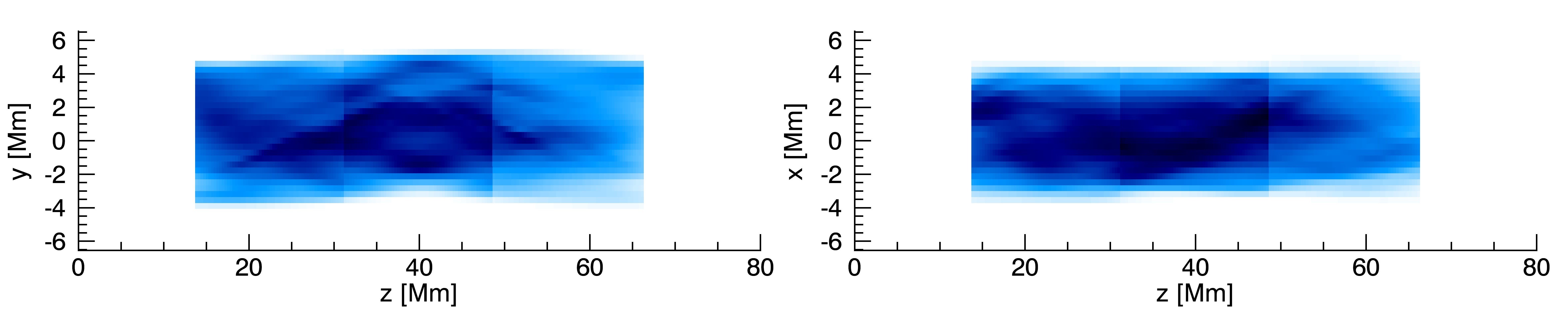} \\
\begin{turn}{90} \hspace{0.8cm} $t=539$\end{turn}& \includegraphics[scale=0.07]{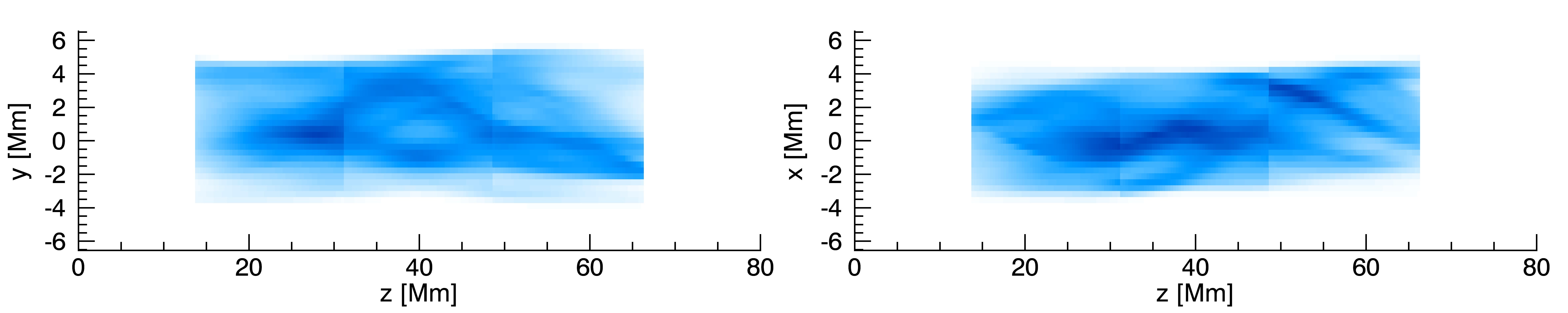} \\
\begin{turn}{90} \hspace{0.8cm} $t=609$\end{turn}& \includegraphics[scale=0.07]{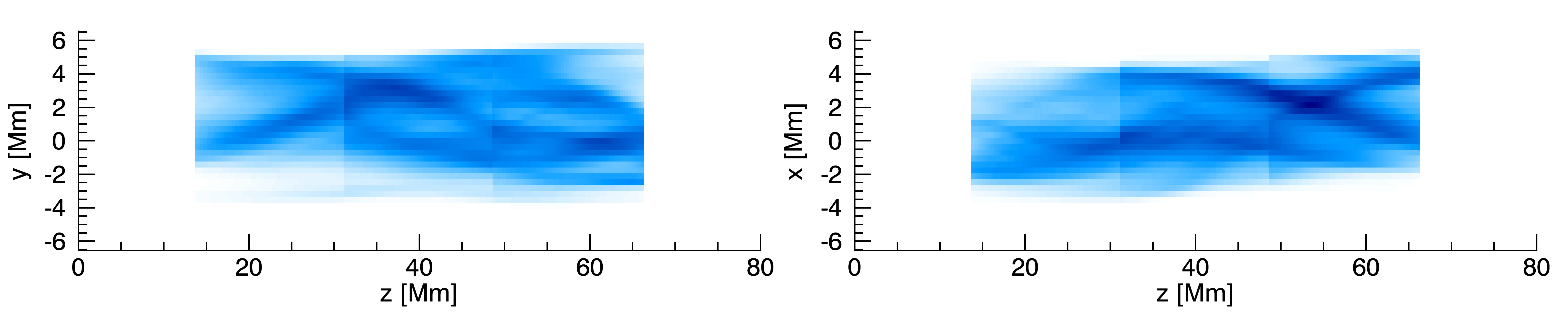} \\
\begin{turn}{90} \hspace{0.8cm} $t=679$\end{turn}& \includegraphics[scale=0.07]{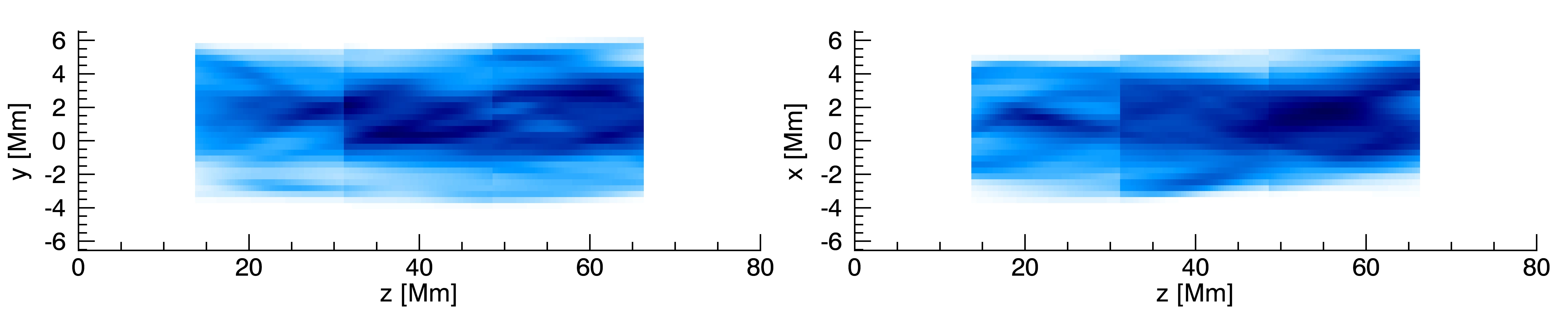}
\end{tabular}
\caption{Intensities Ca \textsc{xv} using a moving mosaic \added{imaging sequence}, integrated along the $x$ (left) and $y$ (right) directions. Dark colour indicates high intensity.}
\label{figmosca15}
\end{figure*}

The three tiles shown in Figure \ref{figtiles} are scanned sequentially from left to right. The resultant intensity mosaics are shown in Figures \ref{figmosfe11}, \ref{figmosfe13} and \ref{figmosca15} for the Fe \textsc{xi}, Fe \textsc{xiii} and Ca \textsc{xv} spectral lines. 

At time $t=261$~s, the current sheets are highly twisted which can be seen in the left panel of all the figures. This twist is short-lived so does not appear across the entire mosaic, only the left tile. Following this, the current sheets reconnect and the peak temperature rises above 12 MK, thereby evolving out of emissivity for the considered spectral channels, resulting in a drop of intensity in all spectral lines. \deleted{There are a few bright points that occur in the centre and right tiles near the radial edge of the loop in the Fe \textsc{xi} intensity maps that are a result of heat being thermally conducted away from reconnection sites.} \added{The loop expands radially outwards and reconnects with the exterior field. Temperature is then thermally conducted along these newly reconnected field lines. The temperature decreases radially outwards from the loop core resulting in successively cooler spectral lines being activated at greater radius. The Fe \textsc{xi} line shows a brightening at the radial edge of the loop as the start of the non-linear phase whereas the Ca \textsc{xv} line has very little emission due to the temperature gradient.}
\added{In the 3D intensity plot (before LOS integration), the loop brightens fairly uniformly on the radial edge. However, when the LOS integration is performed, a LOS through the radial edge of the loop has significantly higher intensity than near the centre since the LOS intercepts more of the bright edge. The bright structures on the radial edge of the intensity map are due to this effect.}

At time $t \approx 400$~s, the average temperature in the loop is sufficient to generate a high intensity in the Ca \textsc{xv} spectral line, revealing the hot interior structure of the loop. The cooler Fe \textsc{xi} spectral line exhibits bright points at the radial edge of the loop as a result of the radial expansion of the loop into the cooler surrounding plasma. 

For time $t > 400$~s, large continuous structures are observable in all three spectral lines and across multiple tiles indicating that these structures inside the loop evolve on longer timescales than the effective cadence of the mosaic, i.e. 69~s. The braided nature of these continuous structures is most visible in the hottest channel here, i.e. Ca~ \textsc{xv}. \added{These braided structures are present in the numerical model \citep[see Figure 7 in][]{Botha2011kink}. DKISL/DL-NIRSP has sufficient spatial and temporal resolution to resolve the braided interior structure in the Ca \textsc{xv}.}

\subsubsection{Doppler velocities}

\begin{figure*}
\vspace{0cm}
\begin{tabular}{c c}
& \hspace{-2cm} x \hspace{8cm} y \\
\begin{turn}{90} \hspace{0.8cm} $t=261$\end{turn}& \includegraphics[scale=0.07]{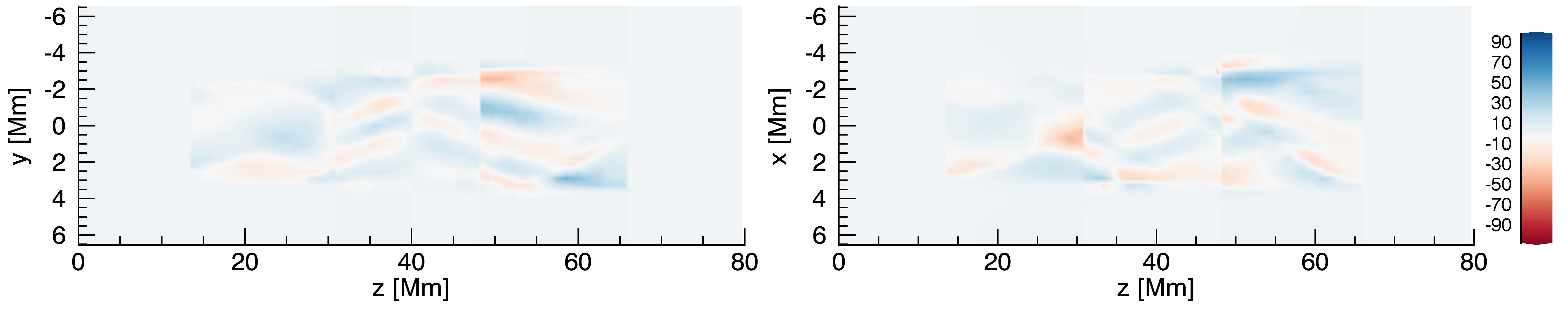} \\
\begin{turn}{90} \hspace{0.8cm} $t=330$\end{turn}& \includegraphics[scale=0.07]{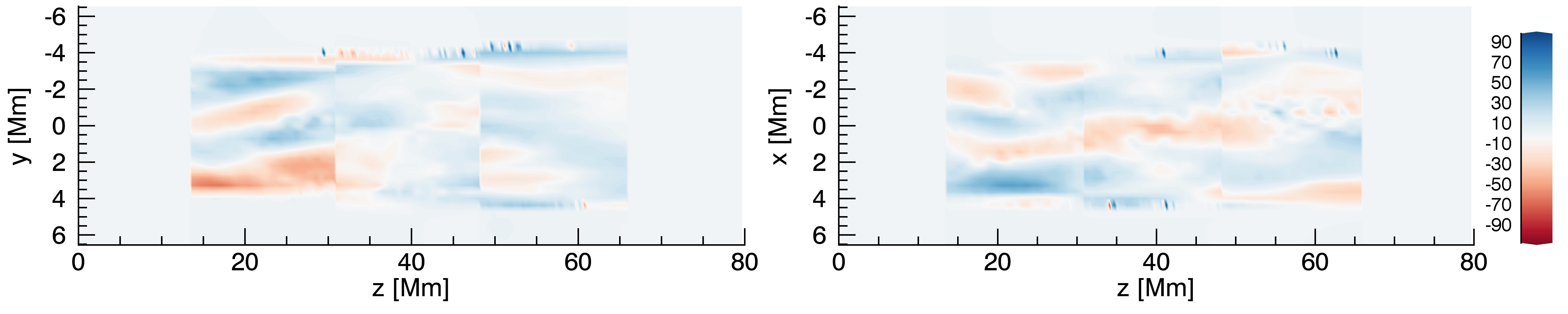} \\
\begin{turn}{90} \hspace{0.8cm} $t=400$\end{turn}& \includegraphics[scale=0.07]{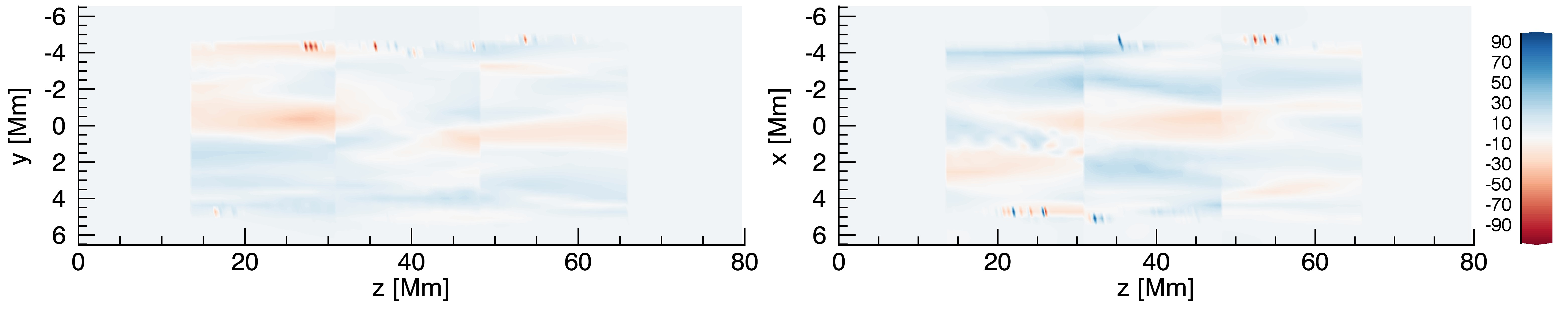} \\
\begin{turn}{90} \hspace{0.8cm} $t=469$\end{turn}& \includegraphics[scale=0.07]{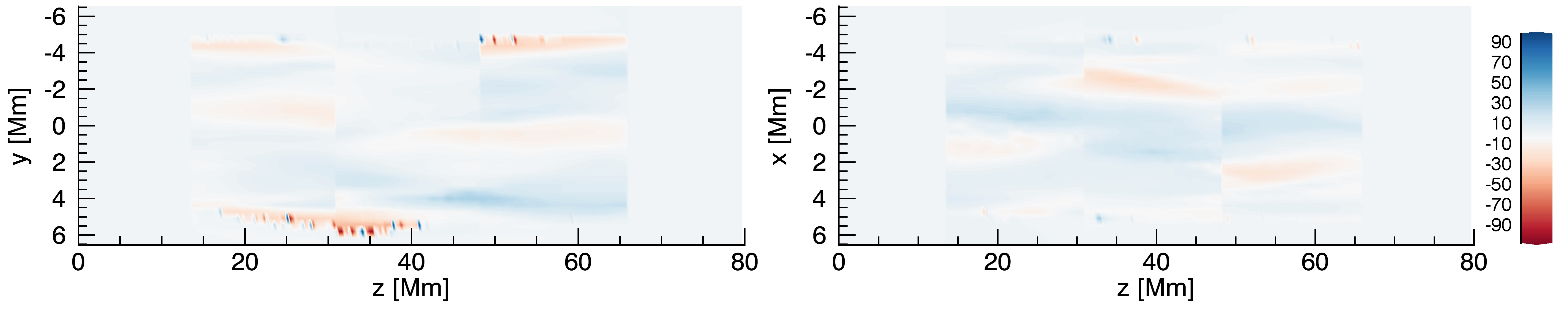} \\
\begin{turn}{90} \hspace{0.8cm} $t=539$\end{turn}& \includegraphics[scale=0.07]{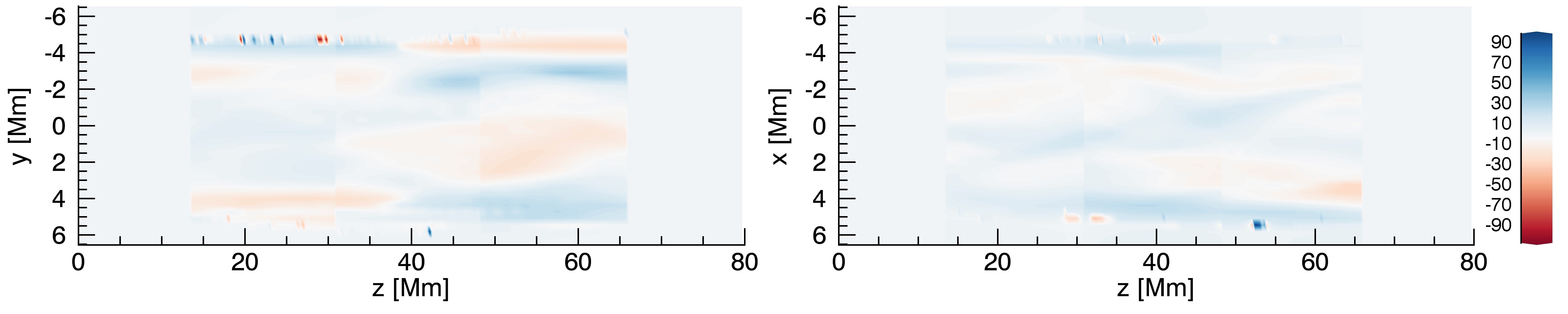} \\
\begin{turn}{90} \hspace{0.8cm} $t=609$\end{turn}& \includegraphics[scale=0.07]{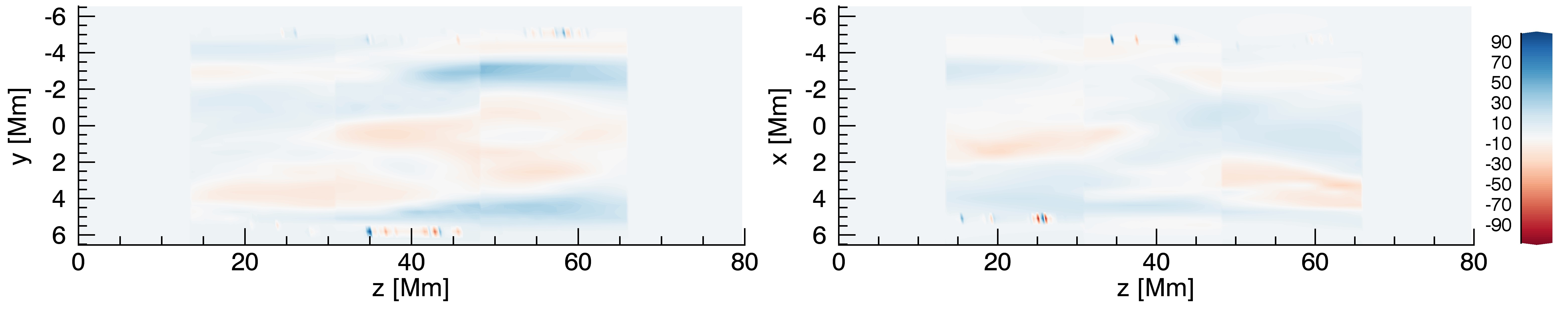} \\
\begin{turn}{90} \hspace{0.8cm} $t=679$\end{turn}& \includegraphics[scale=0.07]{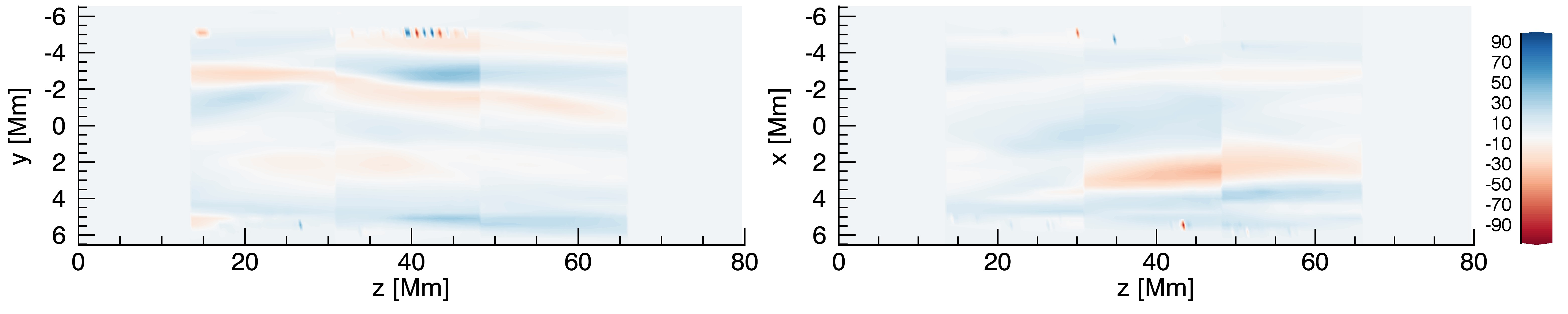}
\end{tabular}
\caption{Doppler velocities using a moving mosaic \added{imaging sequence} and the Fe \textsc{xiii} spectral line. Colour table is between $\pm 100$ km s$^{-1}$.}
\label{figmosfe13dopp}
\end{figure*}

The Doppler velocity mosaics for the Fe \textsc{xiii} spectral line are shown in Figure \ref{figmosfe13dopp}. These are calculated according to the following formula:
\begin{equation}
D= \frac{\int C(T) n_e ^2 \textbf{v} \cdot d\textbf{l}}{\int C(T) n_e ^2 \: dl},
\end{equation}
where $\textbf{v}$ is the velocity and the scalar product with $d\textbf{l}$ produces the Doppler velocity along the LOS. The denominator is the intensity $I$ from Equation (\ref{eqnintens}).

In the Doppler maps, there are a collection of small-scale bursts that can be detected on the radial edge of the loop. These are indicative of small-scale reconnection occurring in the simulation. These are very localised, transient features with velocity magnitudes of up to $\pm 100$ km s$^{-1}$. \added{Magnetic reconnection generates high-velocity flows that are aligned with the magnetic field. The bulk magnetic field is in the $z-$direction along the tube length, however the twisted nature of the magnetic field results in a LOS magnetic field component which guides the velocity, producing these high-velocity bursts in the LOS Doppler maps.}

\subsection{Sit-and-stare}

In sit-and-stare mode, a single tile is observed continuously without scanning in the image plane, hence the cadence is set by the exposure time of 22.9~s per tile, allowing us to investigate the loop sub-structures that evolve on shorter timescales than with the mosaics.

\subsubsection{Intensity}

\begin{figure*}
\vspace{0cm}
\begin{tabular}{c c c c}
\begin{turn}{90} \hspace{0.8cm} $t=261$\end{turn}& \includegraphics[scale=0.07]{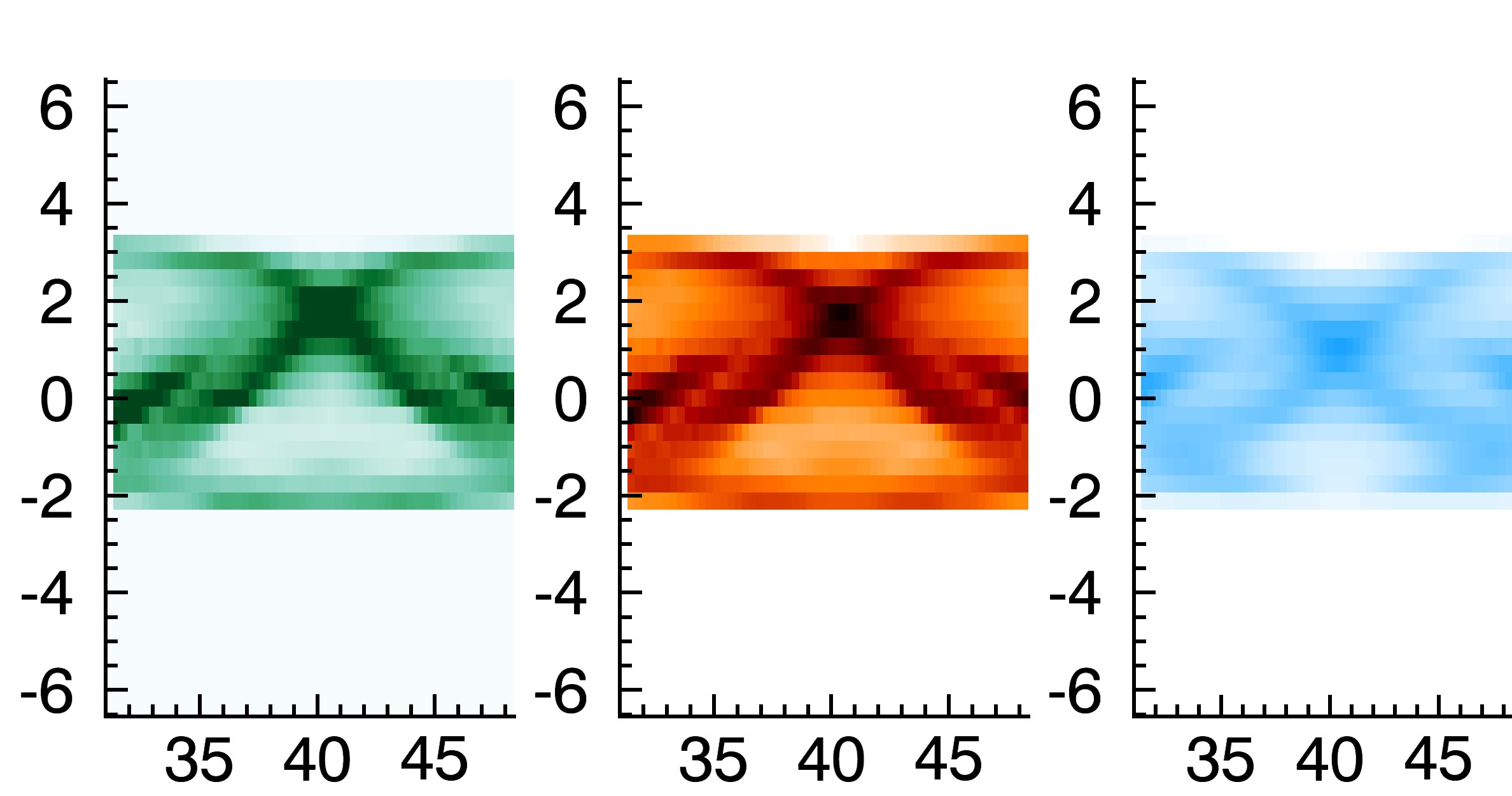} & \begin{turn}{90} \hspace{0.8cm} $t=261$\end{turn}& \includegraphics[scale=0.07]{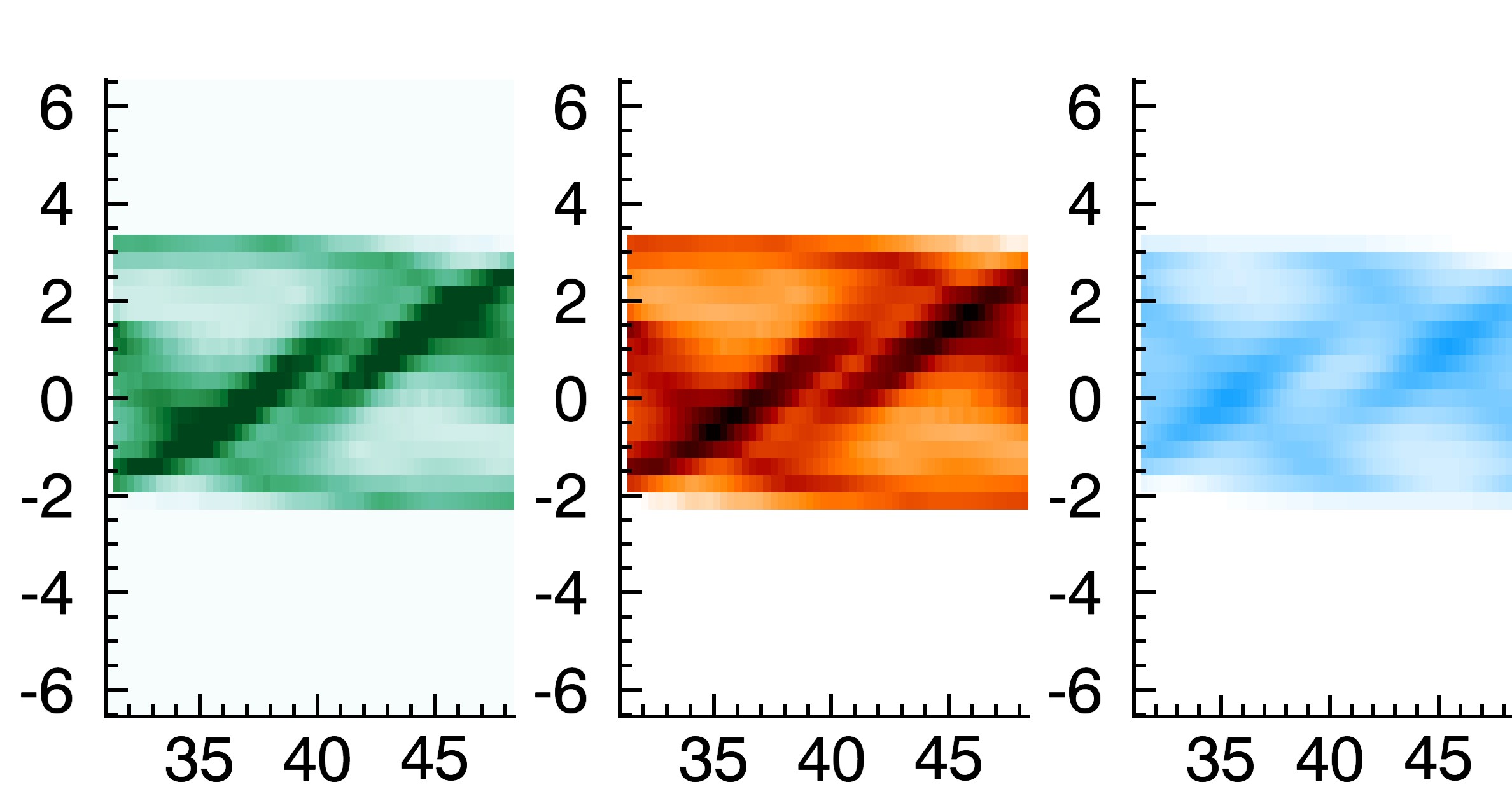}\\
\begin{turn}{90} \hspace{0.8cm} $t=330$\end{turn}& \includegraphics[scale=0.07]{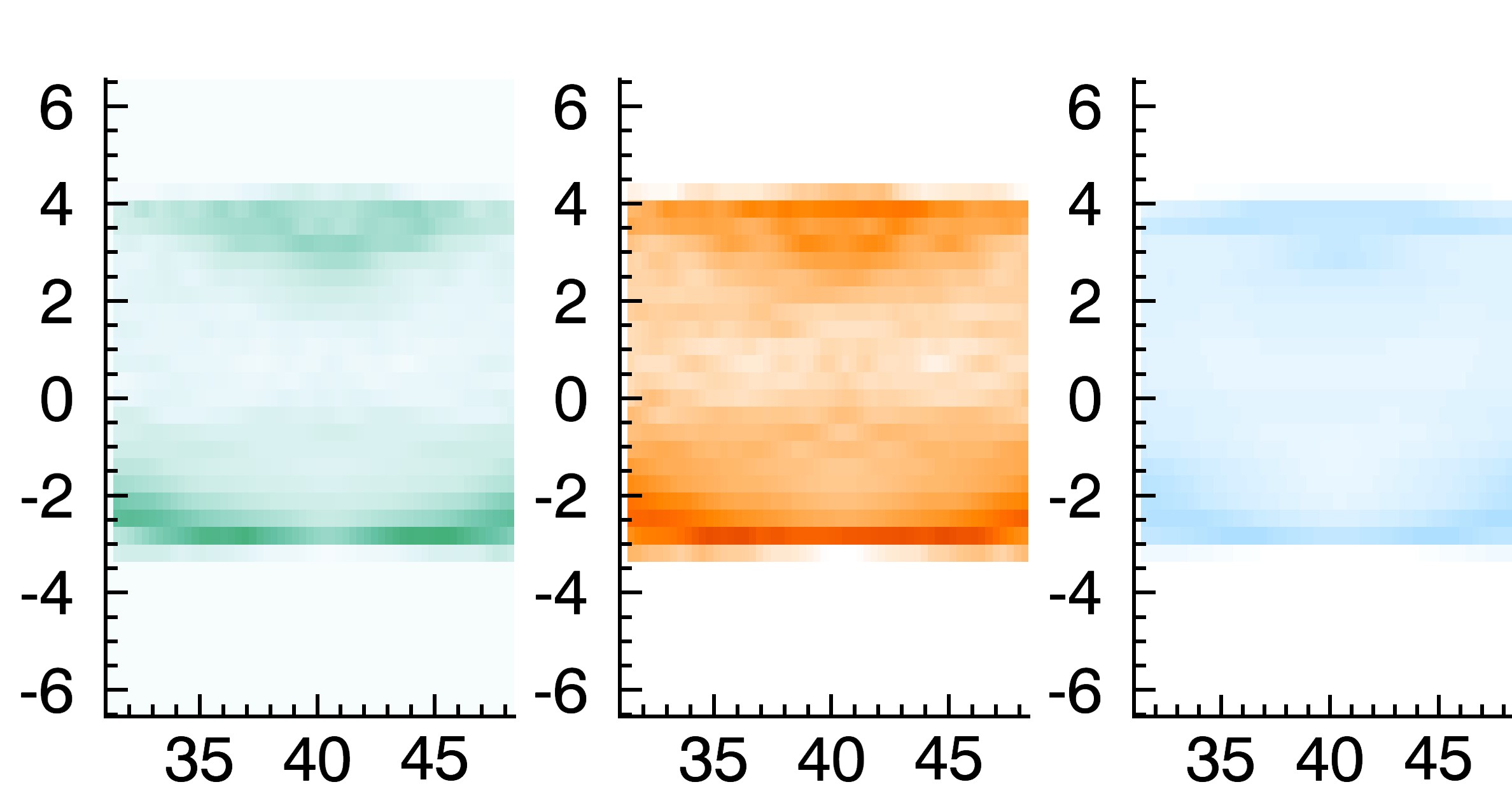} & \begin{turn}{90} \hspace{0.8cm} $t=330$\end{turn}& \includegraphics[scale=0.07]{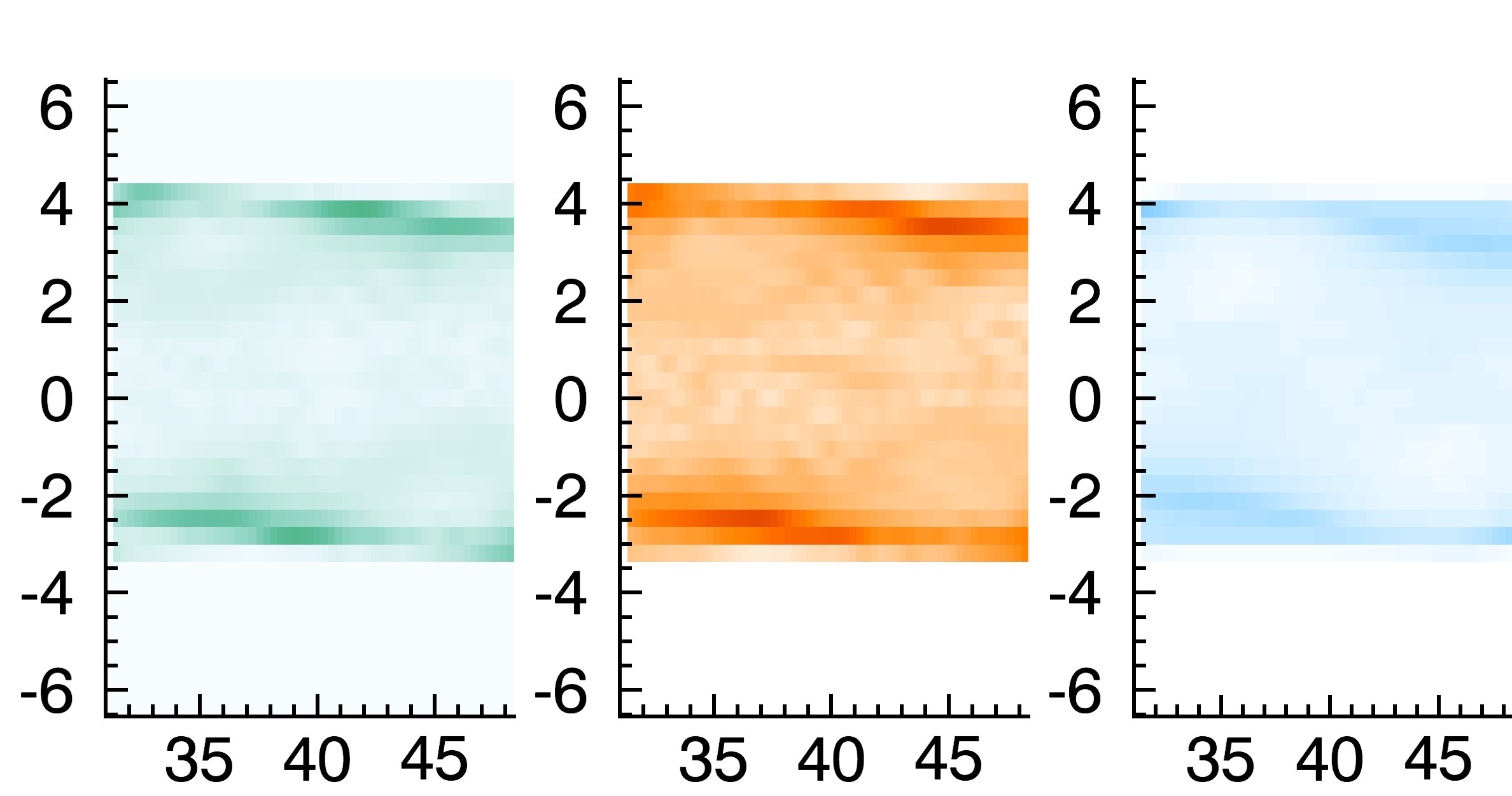}\\
\begin{turn}{90} \hspace{0.8cm} $t=400$\end{turn}& \includegraphics[scale=0.07]{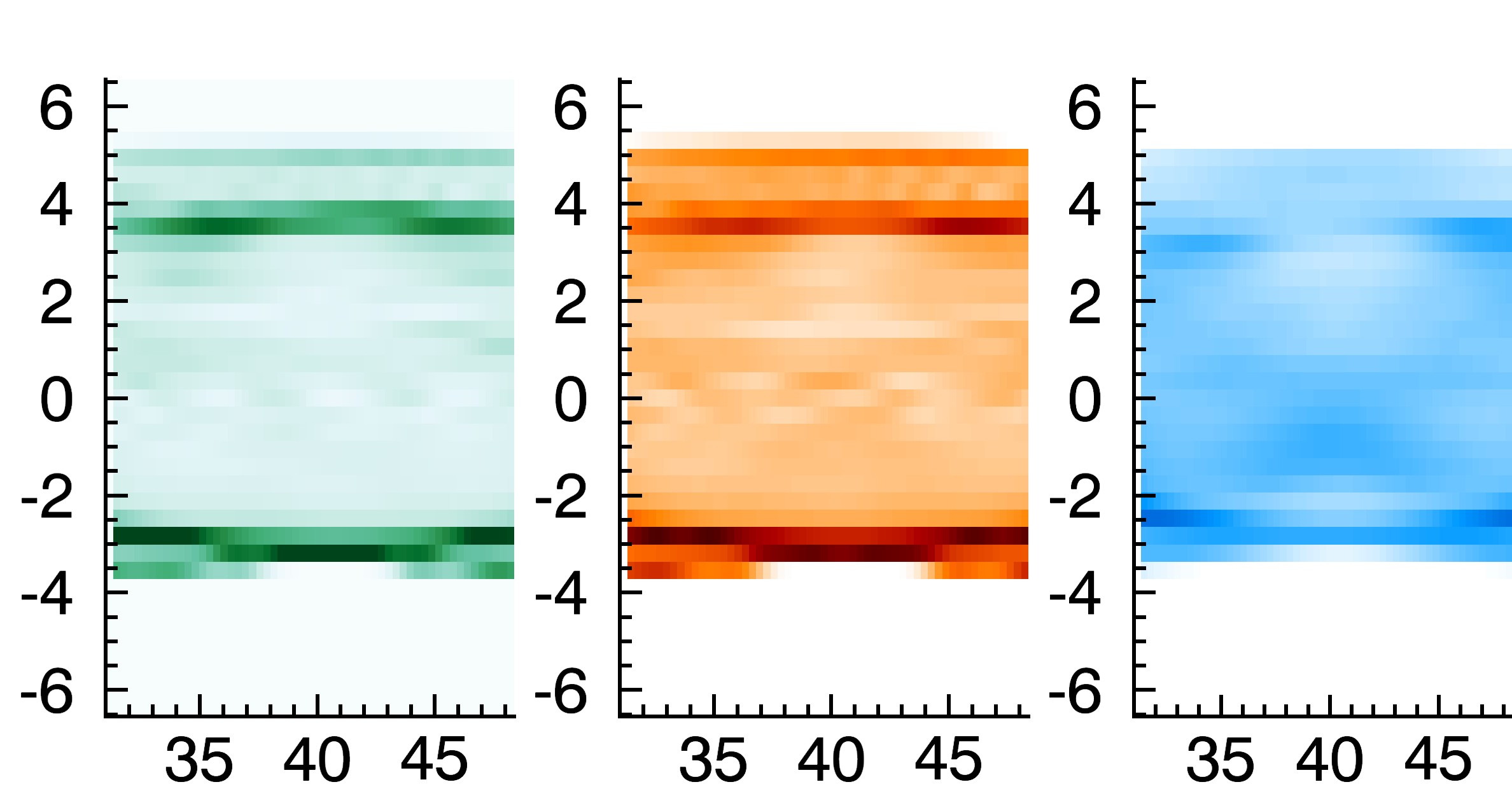} & \begin{turn}{90} \hspace{0.8cm} $t=400$\end{turn}& \includegraphics[scale=0.07]{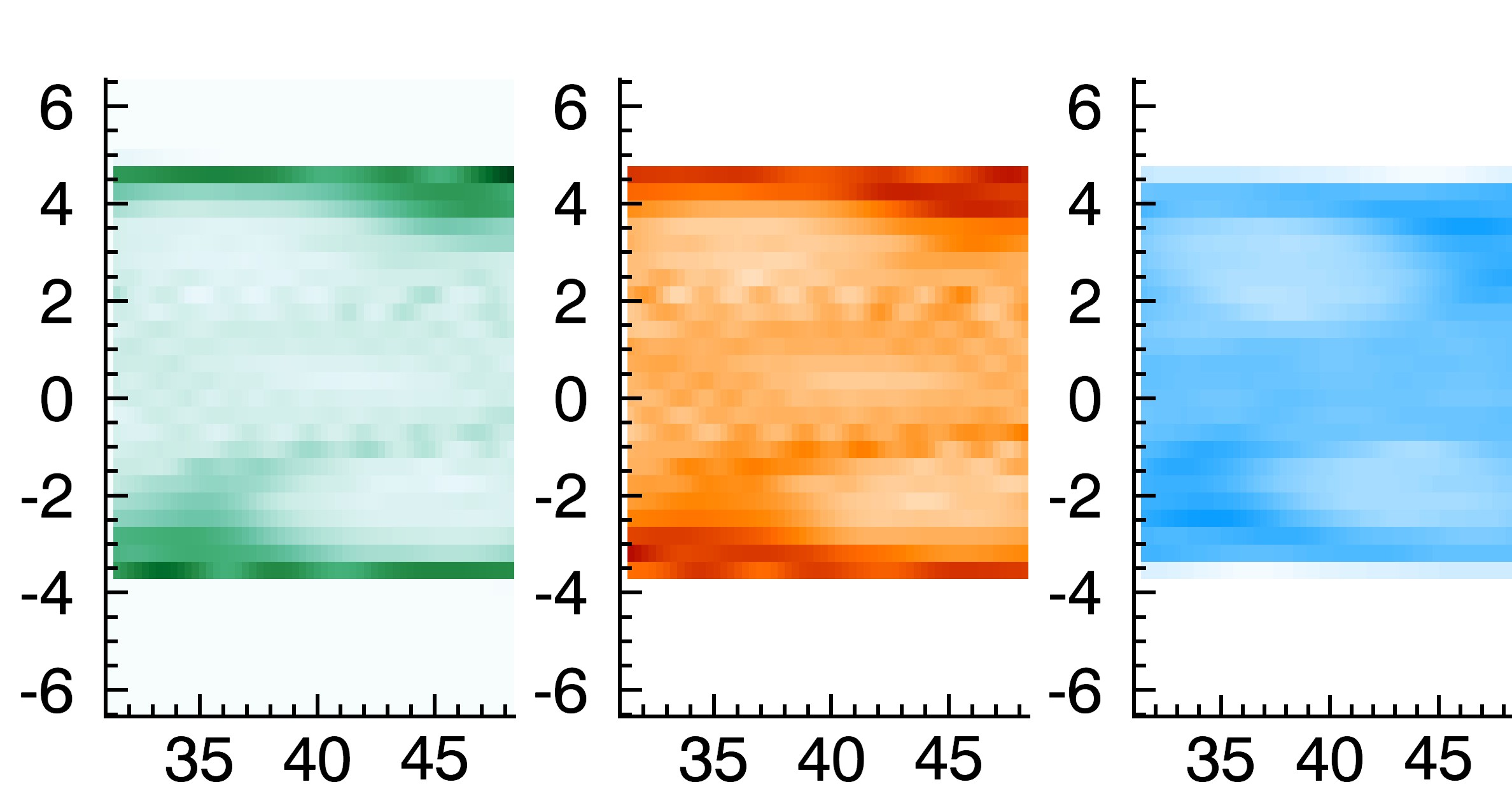}\\
\end{tabular}
\caption{Intensities sit-and-stare using Fe \textsc{xi} (green), Fe \textsc{xiii} (red), and Ca \textsc{xv} (blue) integrated in $x$ (left) and $y$ (right) directions. Mosaic tile is located in the centre of the loop.}
\label{figstare}
\end{figure*}

A few chosen snapshots of the sit-and-stare intensities for the central tile are shown in Figure \ref{figstare} for the three spectral lines. Here we \added{observe clearly} the braided nature of the loop sub-structure early on during the evolution of the kink instability (at $t=261$~s) before the dominant energy transfer changes to the radial loop boundary. For time $t>400$~s the brightest signatures in the coolest spectral channels appear at the loop boundary whereas it remains brightest in the middle of the loop in the Ca~\textsc{xv} line.

\begin{figure*}
\includegraphics[scale=0.33,clip=true, trim=3cm 8cm 3cm 8cm]{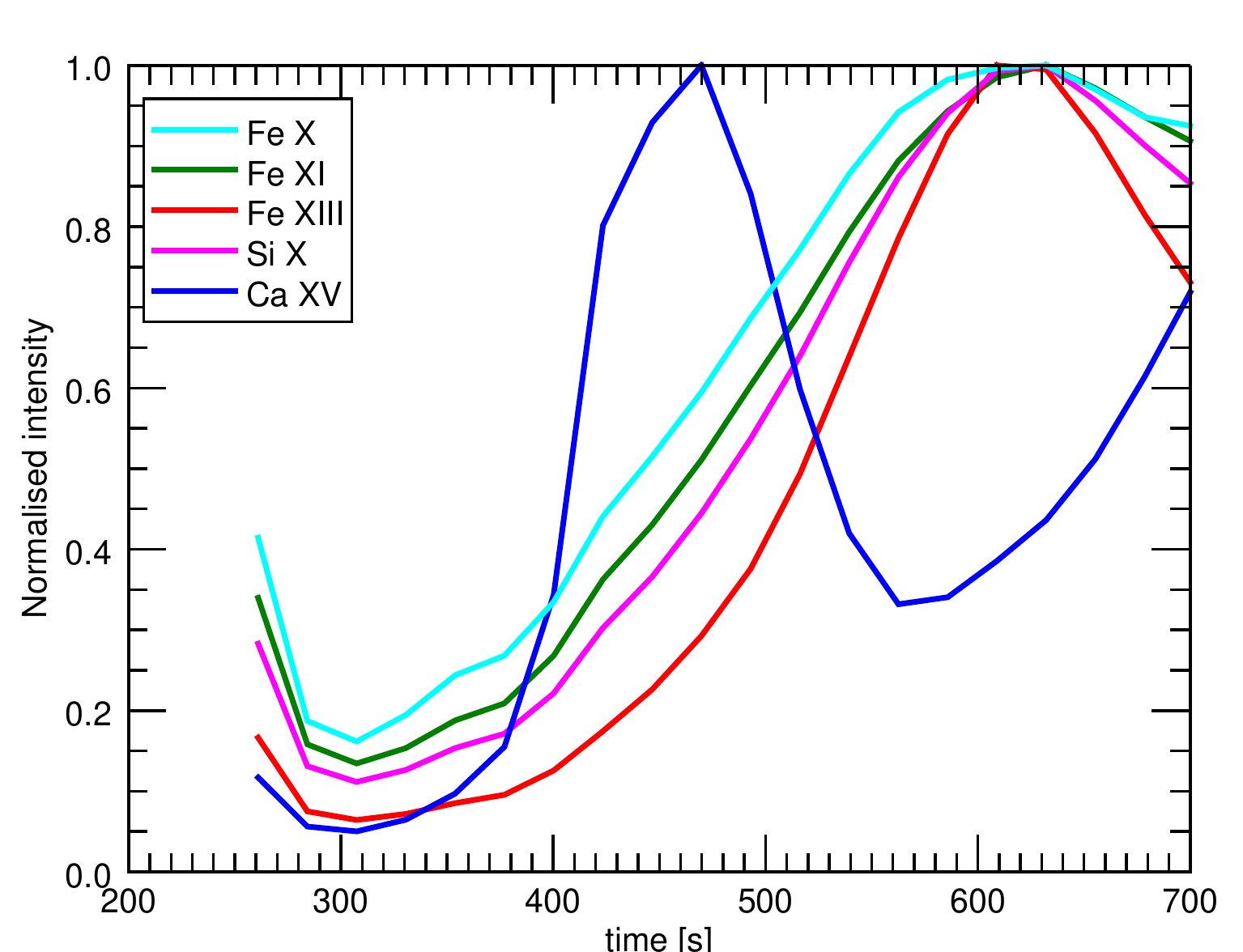}
\includegraphics[scale=0.33,clip=true, trim=3cm 8cm 3cm 8cm]{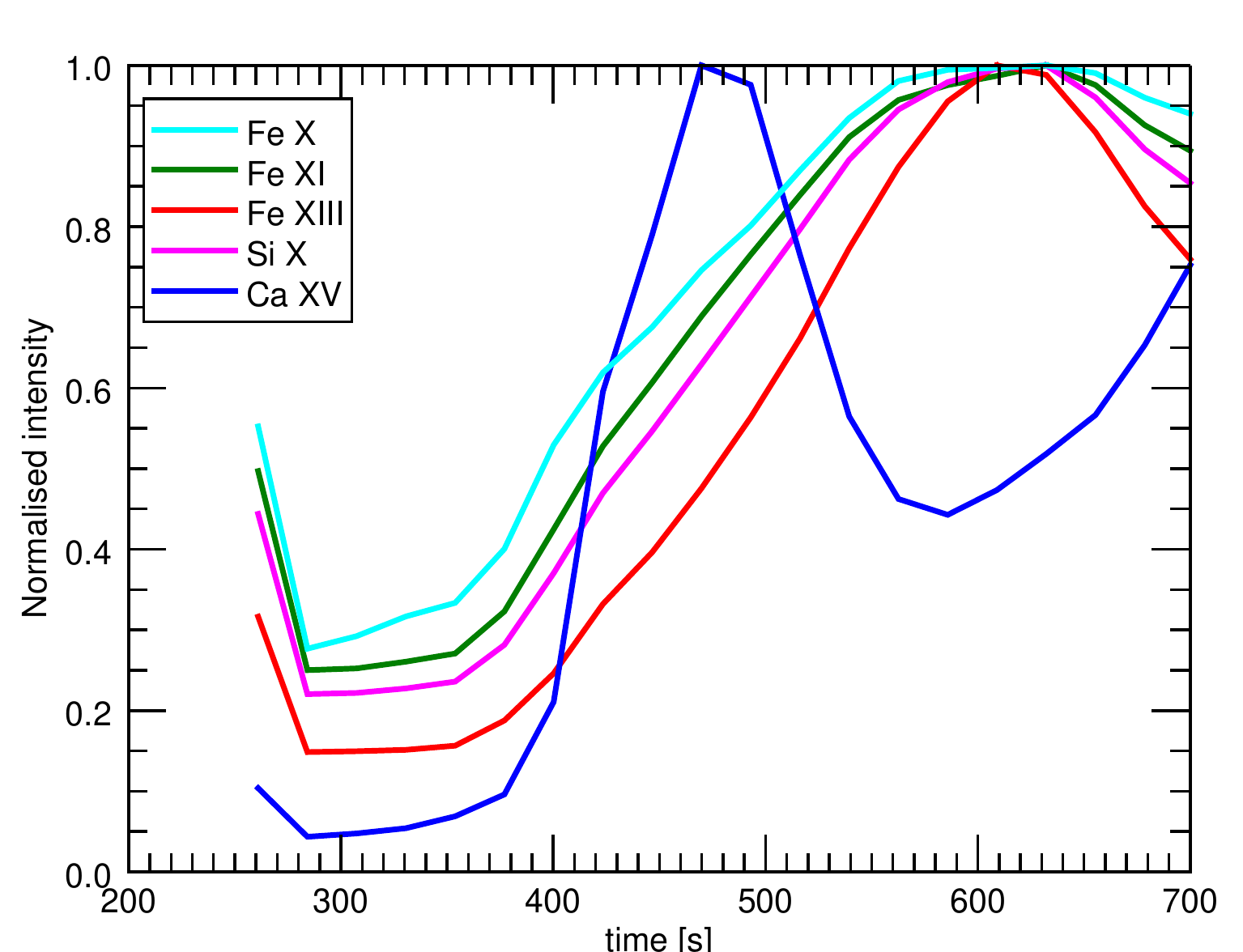}
\includegraphics[scale=0.33,clip=true, trim=3cm 8cm 3cm 8cm]{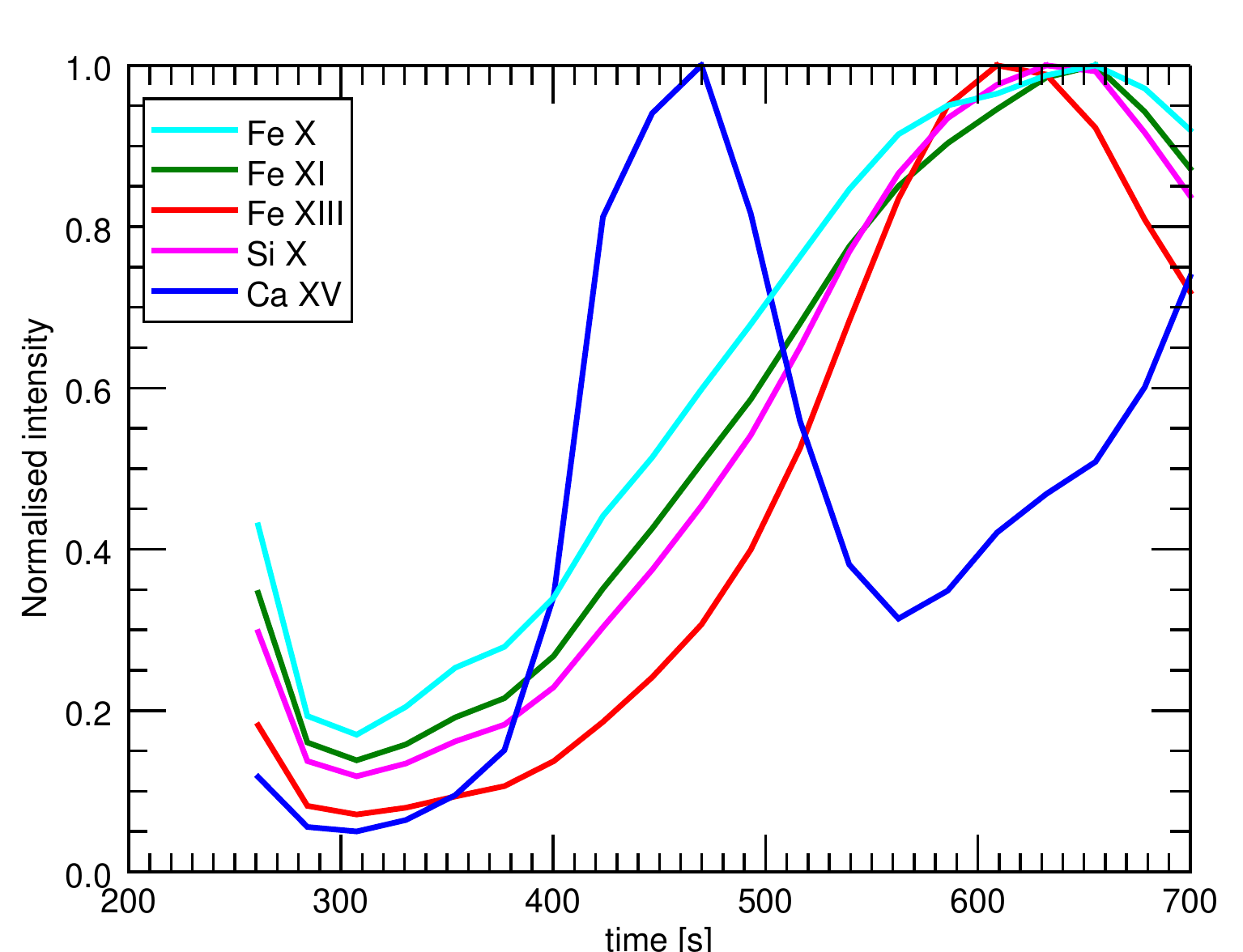}
\caption{Average intensities for sit-and-stare panels located at the left, centre and right of the loop. \added{Normalised to unity at the maximum amplitude.} }
\label{figlight}
\end{figure*}

Light curves of these spectral lines have been synthesised to analyse the energy transport ongoing within the loop from an observational perspective (Figure~\ref{figlight}).
Here we observe that the hottest line of Ca~~\textsc{xv} (blue) reaches peak intensity first, followed then by emission in the cooler lines.
The staggered response in the peak intensities across the different lines demonstrates the temperature inside the loop equalising and activating successively cooler spectral lines.

\begin{figure*}
\includegraphics[scale=0.33,clip=true, trim=3cm 7cm 3cm 8cm]{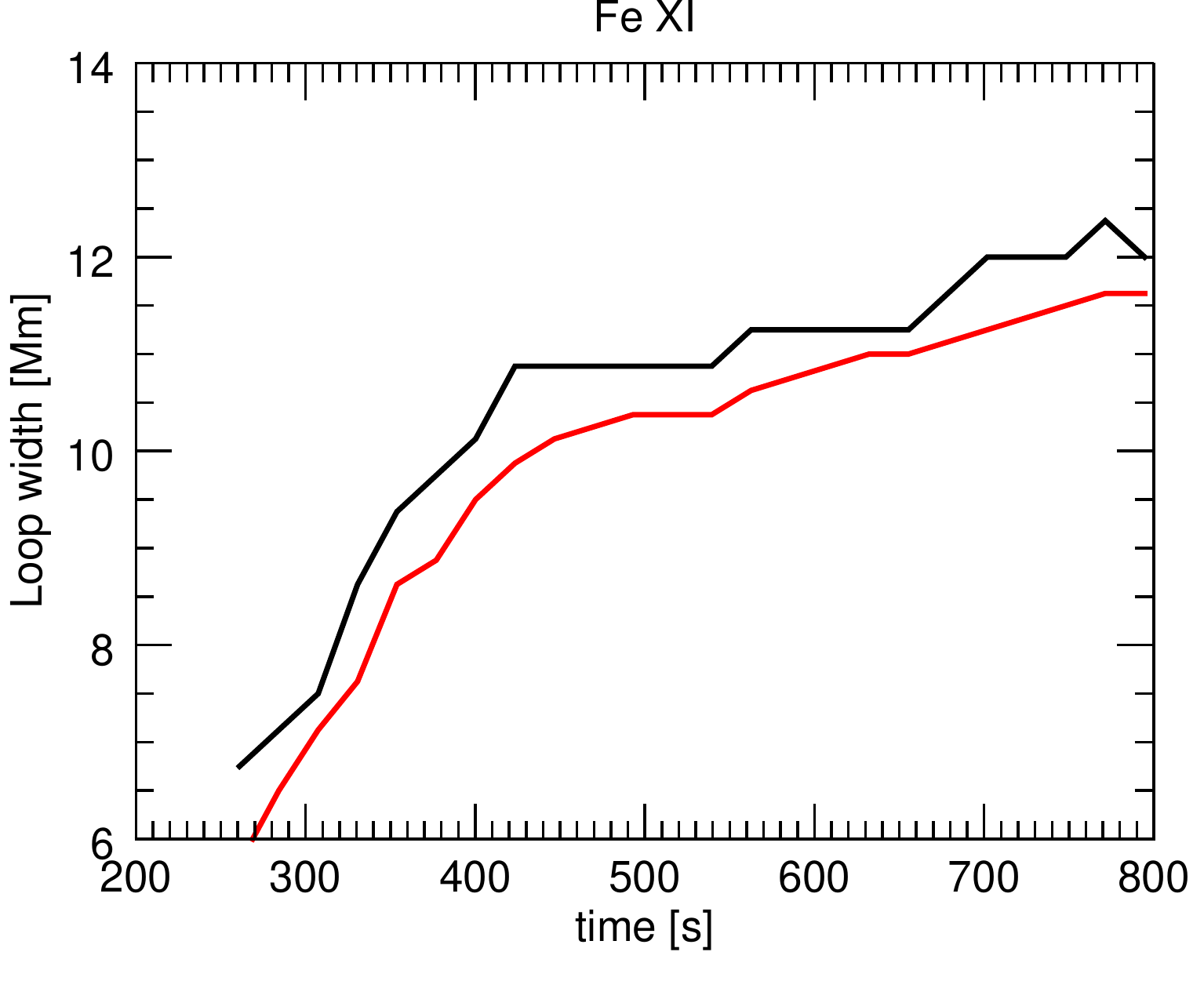}
\includegraphics[scale=0.33,clip=true, trim=3cm 7cm 3cm 8cm]{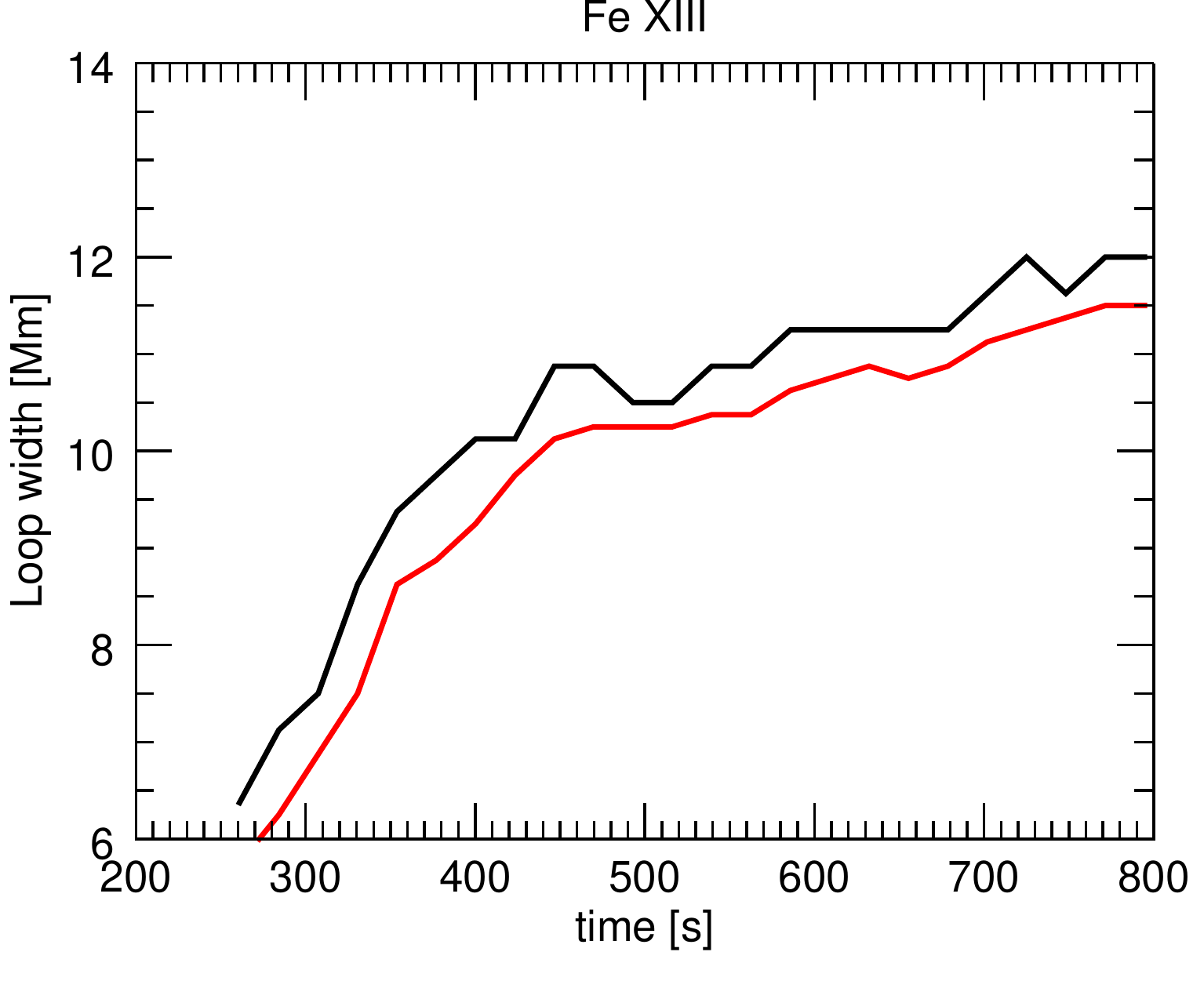} 
\includegraphics[scale=0.33,clip=true, trim=3cm 7cm 3cm 8cm]{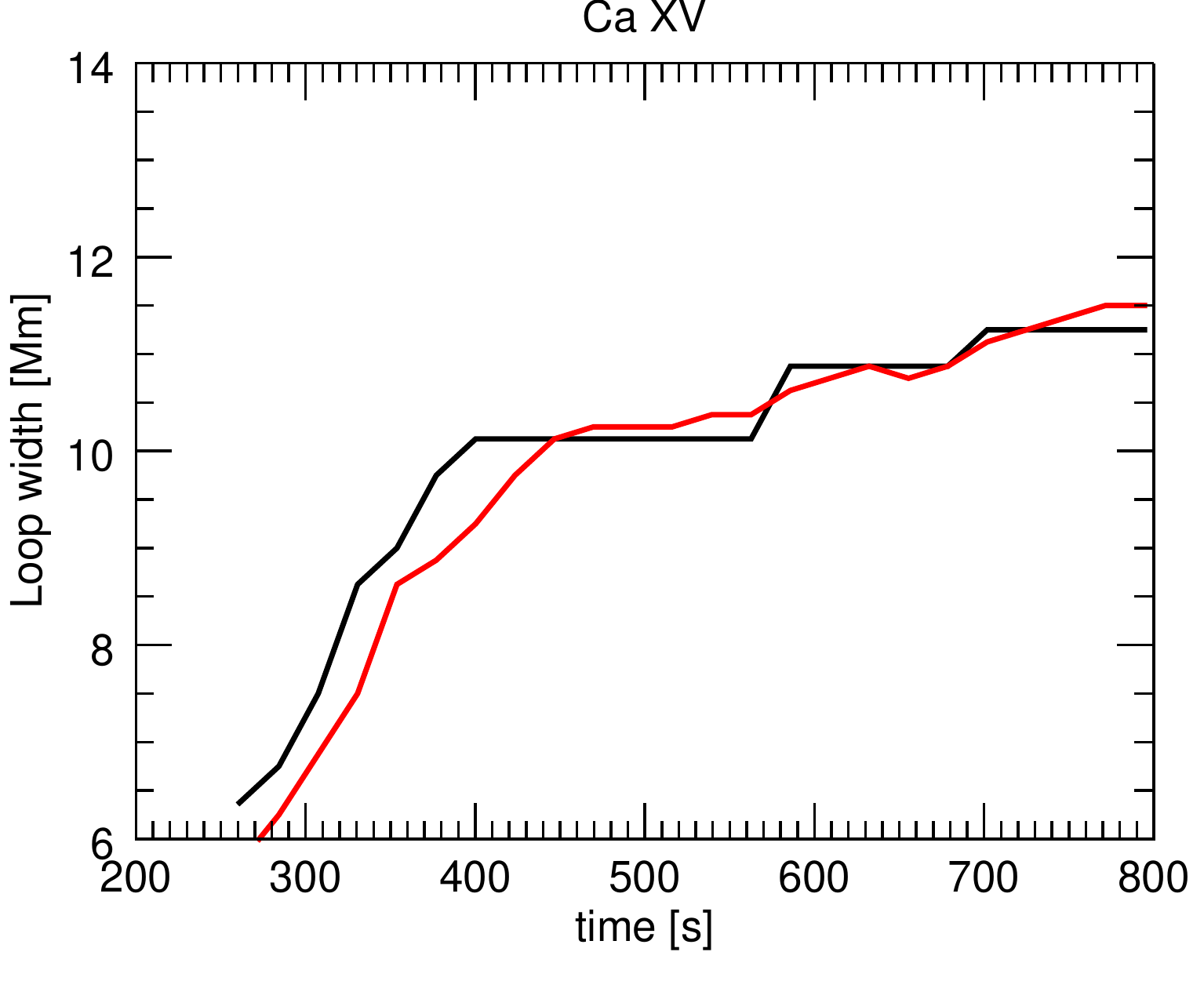}
\caption{Loop width from the sit-and-stare mosaics (black) and the simulation resolution intensities (red) for the Fe \textsc{xi} (left) Fe \textsc{xiii} (centre) and Ca \textsc{xv} (right). The sit-and-stare panel is located at the centre of the loop.}
\label{figloopwidth}
\end{figure*}

The time evolution of the loop width is calculated for the three spectral lines for a mosaic tile located at the centre of the loop, shown in Figure \ref{figloopwidth}. As the non-linear phase of the kink-instability initiates \added{(t=261 s)}, the loop width \added{increases rapidly}, followed by a more gradual increase. The loop width from the synthetic observation is close to the simulation value for all three spectral lines indicating that the loop width can be \added{measured accurately}, despite the spatial and temporal degradation. We see a staggered rise in the loop width, with the loop width increasing first in the cool Fe \textsc{xi} line, followed by the Fe \textsc{xiii} line, and finally the hot Ca \textsc{xv} line. As the loop evolves, the heat is spread radially outwards due to a combination of magnetic reconnection and parallel thermal conduction\added{: reconnection occurs between the loop and the exterior field and thermal conduction transports heat along these newly reconnected field lines} \citep{Botha2011kink}. Therefore, the temperature gradually decreases radially outwards, activating cooler spectral lines. Thus, the cooler Fe \textsc{xi} line shows the radial extent of the loop, whereas the hot Ca \textsc{xv} line shows the loop interior. 

\subsection{Line ratio and density prediction}\label{sec:dens}

\begin{figure}
\centering
\includegraphics[scale=0.45,clip=true, trim=3cm 7.5cm 3cm 8cm]{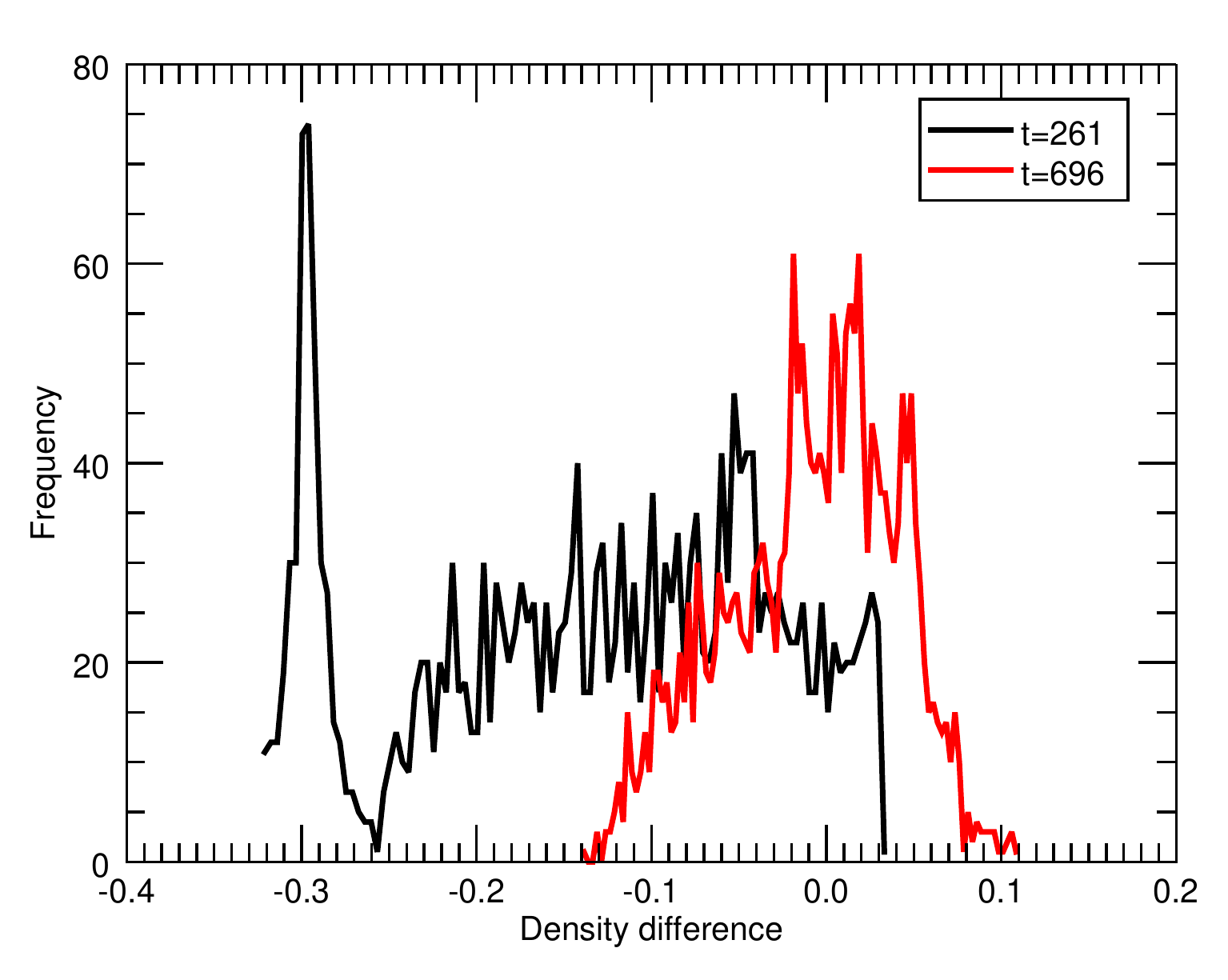}
\caption{Histogram of the difference in predicted density from the Fe \textsc{xiii} line ratios and the mean simulation density along the LOS for every grid-cell in the central tile, scaled by the simulation density. Black line is during a highly dynamic phase ($t=261$ s). Red line is during a less dynamic phase ($t=696$ s).}
\label{figintensratio}
\end{figure}

The ratio of the Fe XIII lines at 10747 and 10798 \deleted{Angstoms}\added{\AA{}} can be used to estimate the electron density of the solar plasma.  The filter for the 10798 \deleted{Angstoms}\added{\AA{}} line may be available in the first light configuration of DKIST/DL-NIRSP, and is a strong candidate for addition in future upgrades.
To test the accuracy of the density estimation for this event, we compare the density estimated from the synthesised ratio of these two lines, to the mean density along the LOS. \added{For this density diagnostic, the contribution function was resynthesised in each numerical cell using the local temperature and density values.} A histogram of the normalised difference is shown in Figure \ref{figintensratio}. The density is estimated at two different times during the evolution. \added{The error is determined by comparing the diagnosed value to the average density along the LOS. The density variations throughout the simulation are reasonably small (much less than an order of magnitude) hence this is a reasonably accurate metric for the density along a LOS.}
The accuracy of the density estimate depends greatly on the density inhomogeneities along the LOS, and whether these fluctuations are in the correct temperature range to activate the Fe \textsc{xiii} lines.

At time $t=261$ s the twisted structure of the loop is present in the Fe \textsc{xiii} line (see Figure \ref{figmosfe13}), meaning that the LOS density inhomogeneities contribute to the intensity, resulting in a peak overestimate of the density by $\approx 30 \%$. 

Late in the simulation ($t \geq 400$ s) the hot core of the loop is highly braided and the LOS density is highly inhomogeneous (as evidenced by the braided structures in Figure \ref{figmosca15}). However, the temperature of the loop core is too hot to fully activate the Fe \textsc{xiii} line, hence the braided interior of the loop is not present in the Fe \textsc{xiii} intensity maps (Figure \ref{figmosfe13}). Instead, the main contributions to the Fe \textsc{xiii} intensity are from the radial edge of the loop (as shown by the high intensity radial features in Figure \ref{figmosfe13}). The density near the radial edge of the loop is far more uniform than the density in the loop centre, and thus the density estimate becomes more accurate, with errors in the range $[-10 \% , +10 \%]$ (see Figure \ref{figintensratio}). We conclude that the density estimate provided by the Fe \textsc{xiii} line ratio is reasonably accurate when features appear stable in the intensity maps.

\subsection{Signal to noise ratio and photon rates} \label{sec:snr}

\begin{figure*}
\includegraphics[scale=0.06,clip=true, trim=0.5cm 20cm 0cm 4cm]{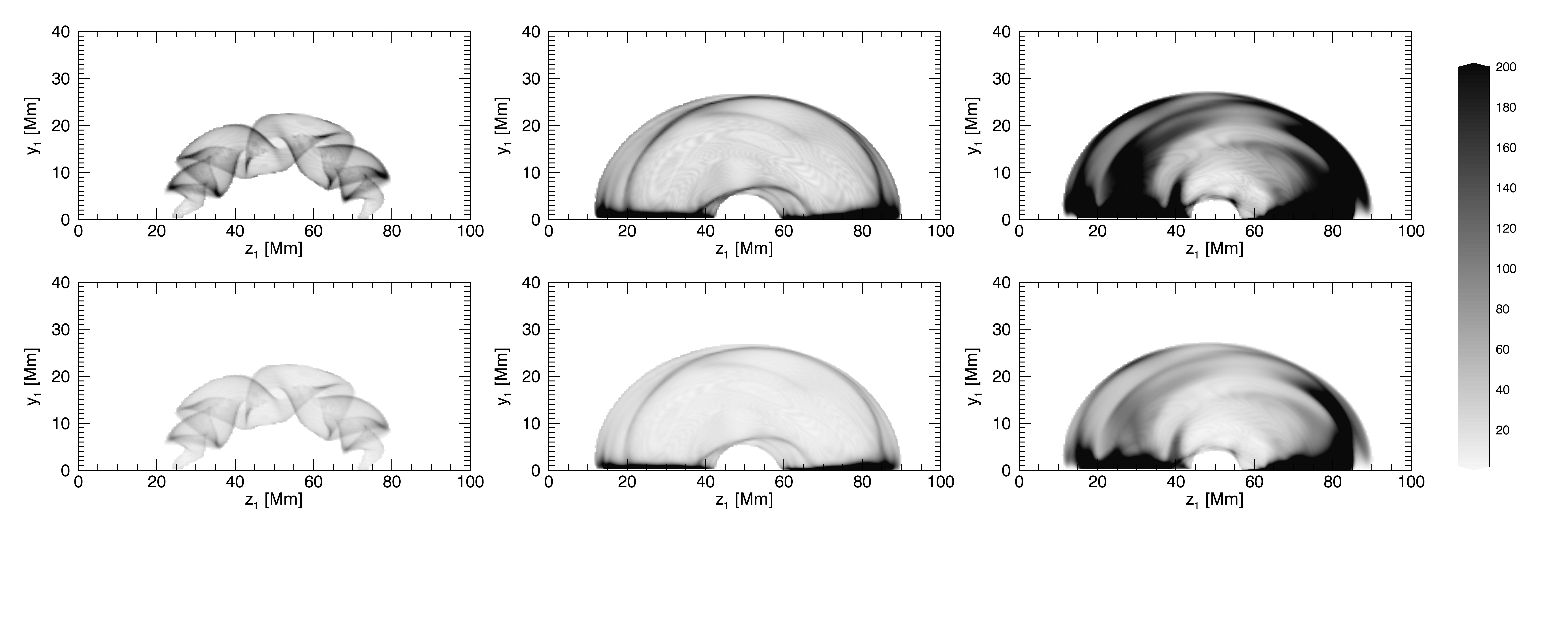}
\caption{Signal-to-noise ratio \added{SNR} in the Fe \textsc{xiii} 10747 spectral line for good (top) and bad (bottom) scattered light conditions at times 261, 469 and 679 seconds from left to right, respectively. All figures have the same colour table and the minimum SNR value is one. $y_1$ and $z_1$ are the new coordinates after the loop is projected to be a curved half-loop with an apex of 25.5 Mm. The SNR is calculated at the line core. \added{The corresponding intensity images of the loop are shown in Figures \ref{figmosfe11}-\ref{figmosca15}}. The SNR effectively mimics the intensity.}
\label{figsnr}
\end{figure*}

The synthetic intensities are used as an input to the DL-NIRSP instrument performance calculator. The instrument calculation uses the telescope area, instrumental efficiency, and instrumental sampling to calculate photons per pixel per second at the detector and adds detector-dependent quantities for read noise and dark current, wavelength and pointing-dependent quantities for the average coronal continuum and sky brightness, and an instrumental scattered light which is a fixed fraction of the disk intensity. This creates a pixel-by-pixel estimation of the \added{SNR} at the peak intensity of the line profile allowing us to test the observability of the results presented in this paper.

The straight loop is projected to be a curved half-loop with an apex of 25.5 Mm. Note that this is purely to simulate the observation and the projection itself does not affect the SNR. Two values of instrumental scattered light are used: good ($2.5 \times 10^{-6}$) and bad ($160 \times 10^{-6}$) allowing us to consider the observability of the loop in a range of scattered light conditions. These estimates for scattered light are based on the expected values for a clean condition (good) and after one week of dust build up (bad).

The total exposure time used in this paper is a combination of the camera exposure and the number of co-adds (i.e., total number of exposures). The camera exposure time has to be selected carefully to avoid overexposure and limit the saturation such that the key features remain observable. The camera exposures are then co-added to increase the total effective exposure. The signal-to-noise scales with the square root of the number of co-adds.

We find that for this event, using a camera exposure time of 8 seconds and three co-adds (i.e., three exposures) produces observable results for both good and bad scattered light conditions in the Fe \textsc{xiii} 10747 spectral line, see Figure \ref{figsnr}. Note that the SNR is calculated at line core only, and this figure does not take temporal resolution or the mosaic FOV into account. In both conditions we obtain a reasonable SNR and the key features of interest are observable, namely the loop expansion, sub-structure and braiding, and energy transport throughout the loop. The footpoints of the loop are overexposed and saturated, however, this region is not of interest in this study and the key dynamics are resolved. Under bad scattered light conditions, the SNR is \added{reduced noticeably} throughout the loop compared to the good scattered light conditions. Many features of the loop remain observable and the SNR is enhanced at the regions of interest, i.e. the loop radial boundary and the interior sub-structure and braiding. Therefore, the observational signatures presented in this paper will be observable by DKIST/DL-NIRSP even using conservative estimates for the scattered light conditions. Note that the SNR plots effectively mimic the intensity since the noise level does not vary much over the domain. As such, the SNR plots act as a qualitative representation of the intensity on the projected axes.

\begin{figure}
\centering
\includegraphics[scale=0.45,clip=true, trim=2cm 7.8cm 3cm 8cm]{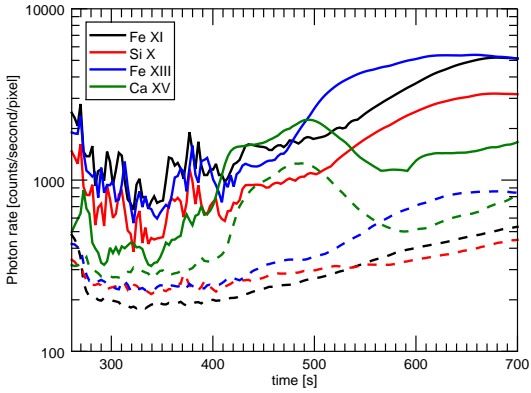}
\caption{Maximum and median photon rate [counts/second/pixel] during the evolution of the loop.}
\label{figphotonrate}
\end{figure}

The other key parameter in determining the observability of the event is the photon rate. The pixel photon rate [counts/second/pixel] is also calculated for the Fe \textsc{xi}, Si \textsc{x}, Fe \textsc{xiii} and Ca \textsc{xv} lines (Figure \ref{figphotonrate}). The median photon count mimics Figure \ref{figlight} and the increase in emission as the loop thermally equalises and activates successively cooler spectral lines. The photon counts per second per pixel is the photon rate multiplied by the camera exposure time, i.e., using a camera exposure of 8 seconds results in a minimum median photon count of $\approx 1600$ counts/s/pixel at the line core of the Fe \textsc{xiii} channel during the low-emission phase of the loop. During the later stages of the instability ($t > 400$ s) the bright edges of the loop have a typical photon count of $\approx 8000$ \added{counts/second/pixel} at the line core. This is far above the noise threshold in both cases and hence the full dynamics of this event will be observable using DKIST/DL-NIRSP. The photon rates are of comparable magnitudes for the other spectral lines indicating that multi-thermal studies of a kink-unstable coronal flux rope will be possible. Note that these photon counts are per detector pixel, and not imaging pixel. The imaging pixel consists of multiple detector pixels and hence the photon count per imaging pixel will be significantly higher. These photon counts are also measured at the line core and could be integrated over the line to further increase the counts.      



\section{Discussion}

Here we have successfully reconstructed DL-NIRSP observables for a kink-unstable coronal loop. These diagnostics will be useful in the interpretation of the observations of such loops at first light, currently projected for 2020. We have also shown that the observed intensity structures are significantly above the noise threshold for the instrument. Next we highlight a number of important physical observables that DL-NIRSP mosaic tiles should detect in identifying such an off-limb loop evolution. 
 
\subsection{Loop sub-structure and braiding}

The large FOV mosaic is capable of detecting transient features and structures present in kink-unstable coronal loops. The initial twist is captured in the first exposure of both the mosaic (Figures \ref{figmosfe11}-\ref{figmosca15}) and sit-and-stare intensity maps (Figure \ref{figstare}). The \added{LOS} Doppler motions can also be measured in these braided sub-structures during this evolution, with motions in the range of $\pm$30~km~s$^{-1}$ in Fe~ \textsc{xiii} near the centre of the loop (Figure \ref{figmosfe13dopp}). As the loop evolves, large structures form along the loop and are identifiable in the moving mosaic intensity maps and these structures are traceable across tiles. Fine structure is also present in these intensity maps, e.g. the braided field lines in Ca \textsc{xv} (Figure \ref{figmosca15}). The Ca \textsc{xv} line shows hot structure in the centre of the loop. Fe \textsc{xi} and Fe \textsc{xiii} show the structure at the edges of the loop. 
\added{This is consistent with previous results for loop oscillations whereby cooler spectral lines are more suitable for observing the edge of the loop \citep[e.g, ][]{Antolin2016}.}

\subsection{Thermodynamic loop expansion}

The expansion (radial increase) of the loop, arising from the thermodynamic changes in the middle of the loop cross-section during the instability, is apparent in both the mosaic and sit-and-stare intensity maps. It is clearest in the sit-and-stare image sequence due to the higher cadence, allowing for more image frames in this early evolution. The loop width in the sit-and-stare intensity maps agrees closely with the simulation loop. The radial edge can also be estimated from these sit-and-stare intensities, as presented in Figure \ref{figloopwidth}. The observed loop width (black line) is very close to the loop width of the simulation resolution intensities (red line). However, there remains a notable offset in the full simulation width measurement versus the synthetic observation loop width in particular for relatively cooler lines, whereas the hottest lines show a better agreement with the loop edge. There is also an ordering in the spectral line formation in time, whereby the cooler Fe \textsc{xi} spectral line increases in width first, followed by the Fe \textsc{xiii} and finally the hot Ca \textsc{xv}. The discussion surrounding the location of the edge of a coronal loop is important in the solar physics literature since it dictates how we understand energy deposition across loop boundaries for coronal heating.

\subsection{Small-scale loop boundary bursts}

Small-scale bursts present in the Doppler velocity maps are indicative of small-scale reconnection events at the loop radial boundary as the loop expands during its evolution into the background medium. Most interesting is the observable bi-directional nature of the flows in these small-scale bursts and we detect adjacent Doppler motions on the order of $\pm$100~km~s$^{-1}$ at the loop radius. These occur most frequently in the Fe \textsc{xiii} channel (Figure \ref{figmosfe13dopp}) but also occur in the other channels, albeit less frequently. \deleted{In the relatively cooler channels these small-scale, localised, frequent events at the radial loop edge may be contributing to the net intensity there which is detectable in the sit-and-stare image sequences. }
Furthermore, if this is indeed small-scale magnetic reconnection it explains the redistribution of the energy to beyond the loop edge according to the full simulation resulting in a larger loop width in the synthetic observable (black line) than that of the simulation width (red line), as shown in Figure~\ref{figloopwidth}. 
Whilst at the same time, the Doppler velocity bursts are less frequent in the hotter channels (such as Ca~\textsc{xv})  and the emission in this channel is largely concentrated (from imaging) within the middle of the loop cross-section. 
\deleted{This correspondence implies that the Ca~\textsc{xv} signatures should correspond to structures present within the loop edge making it a better marker for the true location of the loop edge. Indeed, this point is demonstrated in Figure~\ref{figloopwidth} for Ca~\textsc{xv} where we show a much closer correspondence between the location of the observed edge (black line) and the edge according to the full simulation (red line).} 
The ordering of the cooler line responses in the measurement of the loop edge can also be explained by the evolution of the instability, i.e., sudden loop expansion should drive a larger burst rate at the loop edge early on, followed by a reduced burst rate later as the loop system restores itself to a new equilibrium. Aside from the effects due to radial expansion of the loop, we also detect the presence of flows along the axis of the loop that evolve according to the thermal conduction timescale.

\subsection{Energy transport through the loop}

From the light curves of Figure~\ref{figlight} we can detect a sequential ordering of responses in spectral lines evolving from hot lines that respond first, followed by cooler lines that respond later. This strongly indicates cooling processes in the loop after the onset of the instability leading to a redistribution of heat along the loop axis. That redistribution of heat is present in the form of large scale motions (flows) transporting energy away from the central tile in the mosaic. We have detected this signature of energy transport in the measurement of the responses of the spectral lines between the left and right tiles relative to the centre tile. The left and right panels of the loop increase in intensity before the central tile. This shows the loop heating away from the centre of the loop along the axis of the loop, and this is then followed by heat conducting back towards the centre again along the loop axis. Mass motions are also considered to take place during this redistribution through measurements of changes in the density along the axis, which we have calculated through taking the synthetic Fe \textsc{xiii} line ratios from the intensity maps.

\subsection{Density diagnostics}

Fe \textsc{xiii} line ratios provide a good estimate of the density throughout the loop. Through integrating density along a ray passing through the simulated loop we calculated the line intensities for the line ratio pair using the CHIANTI contribution function. From the line ratio maps we determined the profile for the line ratio as a function of density. In this simulation, we could compare how closely matched the line ratios are in determining the density distribution along the loop with that for an isothermal atmosphere derived through CHIANTI.
The estimated density is fairly accurate ($\pm 10 \%$) during the later stages of the instability where the system is evolving slowly in the Fe \textsc{xiii} intensity channel. During the initial impulsive stage the estimated density becomes less accurate and overpredicts the density by $\approx 30 \%$ due to contributions from LOS density inhomogeneities. Therefore, it will be possible to estimate the density using the Fe \textsc{xiii} pair but the accuracy improves greatly when features remain reasonably stable in the Fe \textsc{xiii} intensity maps.

\added{ \subsection{Comparison to existing instruments} Previous forward modelling studies have been performed investigating the observational signatures of a kink-unstable coronal loop using EUV \citep{Botha2012,Snow2017} and SXR/HXR \citep{Pinto2016} instruments. The study presented in this paper demonstrates that DL-NIRSP is a significant advancement on current instrumentation and is able to resolve new signatures of energy release and transport in kink-unstable flux ropes, namely, the high spatial and temporal cadence of DL-NIRSP captures dynamic, small-scale events such as magnetic reconnection, interior braiding, and accurate measurement of LOS density. However, the lines used by the coronal mode of DL-NIRSP cover a fairly small temperature range which misses the temperature extremes that could be measured using HXR or EUV lines. As such, whilst DL-NIRSP represents a significant advance, it would have to be used in conjunction with existing instruments to capture the full thermal structure of the loop.}

\subsection{Summary}

The off-limb coronal mode of the forthcoming DKIST/DL-NIRSP instrument is capable of observing many signatures of energy release in a kink-unstable coronal loop. Notably, we observe the loop sub-structure, radial growth, small-scale Doppler bursts, and thermal structure of the loop. 
We are also able to estimate the LOS density using \added{the} Fe \textsc{xiii} pair to a reasonable degree of accuracy \added{($\pm 10 \%$)}. 
The signatures demonstrate that the forthcoming DKIST/DL-NIRSP instrument will provide a significant advance on current observations and will be able to provide revolutionary observations to help understand the release of magnetic energy in the solar corona. 

\section*{Acknowledgements}
CHIANTI is a collaborative project involving George Mason University, the University of Michigan (USA) and the University of Cambridge (UK). BS was supported by the STFC grant ST/M000826/1. PRY acknowledges funding from NASA grant NNX15AF25G. GJJB, ES and JAM acknowledge STFC for IDL support as well as support via ST/L006243/1.


\bibliographystyle{aasjournal} 
\bibliography{losbib} 

\begin{thebibliography}{}
\expandafter\ifx\csname natexlab\endcsname\relax\def\natexlab#1{#1}\fi

\bibitem[{{Antolin} {et~al.}(2016){Antolin}, {De Moortel}, {Van Doorsselaere},
  \& {Yokoyama}}]{Antolin2016}
{Antolin}, P., {De Moortel}, I., {Van Doorsselaere}, T., \& {Yokoyama}, T.
  2016, \apjl, 830, L22

\bibitem[{Arber {et~al.}(2001)Arber, Longbottom, Gerrard, \& Milne}]{Arber2001}
Arber, T., Longbottom, A.~W., Gerrard, C., \& Milne, A.~M. 2001, J. Comput.
  Phys, 171

\bibitem[{{Bareford} {et~al.}(2016){Bareford}, {Gordovskyy}, {Browning}, \&
  {Hood}}]{Bareford2016}
{Bareford}, M.~R., {Gordovskyy}, M., {Browning}, P.~K., \& {Hood}, A.~W. 2016,
  \solphys, 291, 187

\bibitem[{{Bareford} \& {Hood}(2015)}]{Bareford2015}
{Bareford}, M.~R., \& {Hood}, A.~W. 2015, Philosophical Transactions of the
  Royal Society of London Series A, 373, 20140266

\bibitem[{{Botha} {et~al.}(2011){Botha}, {Arber}, \& {Hood}}]{Botha2011kink}
{Botha}, G.~J.~J., {Arber}, T.~D., \& {Hood}, A.~W. 2011, A\&\ignorespaces A,
  525, A96

\bibitem[{{Botha} {et~al.}(2012){Botha}, {Arber}, \& {Srivastava}}]{Botha2012}
{Botha}, G.~J.~J., {Arber}, T.~D., \& {Srivastava}, A.~K. 2012, \apj, 745, 53

\bibitem[{{Browning} {et~al.}(2008){Browning}, {Gerrard}, {Hood}, {Kevis}, \&
  {van der Linden}}]{Browning2008}
{Browning}, P.~K., {Gerrard}, C., {Hood}, A.~W., {Kevis}, R., \& {van der
  Linden}, R.~A.~M. 2008, \aap, 485, 837

\bibitem[{{Cheung} \& {Isobe}(2014)}]{Cheung2014}
{Cheung}, M.~C.~M., \& {Isobe}, H. 2014, Living Reviews in Solar Physics, 11, 3

\bibitem[{{De Moortel} {et~al.}(2015){De Moortel}, {Antolin}, \& {Van
  Doorsselaere}}]{Demoortel2015}
{De Moortel}, I., {Antolin}, P., \& {Van Doorsselaere}, T. 2015, \solphys, 290,
  399

\bibitem[{{Del Zanna} {et~al.}(2015){Del Zanna}, {Dere}, {Young}, {Landi}, \&
  {Mason}}]{DelZanna2015}
{Del Zanna}, G., {Dere}, K.~P., {Young}, P.~R., {Landi}, E., \& {Mason}, H.~E.
  2015, \aap, 582, A56

\bibitem[{Dere {et~al.}(1997)Dere, Landi, Mason, Fossi, \&
  Young}]{dere1997chianti}
Dere, K., Landi, E., Mason, H., Fossi, B.~M., \& Young, P. 1997, Astronomy and
  Astrophysics Supplement Series, 125, 149

\bibitem[{{Gordovskyy} {et~al.}(2016){Gordovskyy}, {Kontar}, \&
  {Browning}}]{Gordovskyy2016}
{Gordovskyy}, M., {Kontar}, E.~P., \& {Browning}, P.~K. 2016, \aap, 589, A104

\bibitem[{{Hood} {et~al.}(2009){Hood}, {Browning}, \& {van der
  Linden}}]{Hood2009}
{Hood}, A.~W., {Browning}, P.~K., \& {van der Linden}, R.~A.~M. 2009,
  A\&\ignorespaces A, 506, 913

\bibitem[{{Ishii} {et~al.}(1998){Ishii}, {Kurokawa}, \& {Takeuchi}}]{Ishii1998}
{Ishii}, T.~T., {Kurokawa}, H., \& {Takeuchi}, T.~T. 1998, \apj, 499, 898

\bibitem[{{Kumar} {et~al.}(2017){Kumar}, {Yurchyshyn}, {Cho}, \&
  {Wang}}]{Kumar2017}
{Kumar}, P., {Yurchyshyn}, V., {Cho}, K.-S., \& {Wang}, H. 2017, \aap, 603, A36

\bibitem[{{Mandal} {et~al.}(2016){Mandal}, {Magyar}, {Yuan}, {Van
  Doorsselaere}, \& {Banerjee}}]{Mandal2016}
{Mandal}, S., {Magyar}, N., {Yuan}, D., {Van Doorsselaere}, T., \& {Banerjee},
  D. 2016, \apj, 820, 13

\bibitem[{{Miki\'c} {et~al.}(1990){Miki\'c}, {Schnack}, \& {van
  Hoven}}]{Mikic1990}
{Miki\'c}, Z., {Schnack}, D.~D., \& {van Hoven}, G. 1990, \apj, 361, 690

\bibitem[{{Pariat} {et~al.}(2015){Pariat}, {Dalmasse}, {DeVore}, {Antiochos},
  \& {Karpen}}]{Pariat2015}
{Pariat}, E., {Dalmasse}, K., {DeVore}, C.~R., {Antiochos}, S.~K., \& {Karpen},
  J.~T. 2015, \aap, 573, A130

\bibitem[{{Parker}(1988)}]{Parker1988}
{Parker}, E.~N. 1988, \apj, 330, 474

\bibitem[{{Parnell} \& {De Moortel}(2012)}]{Parnell2012}
{Parnell}, C.~E., \& {De Moortel}, I. 2012, Philosophical Transactions of the
  Royal Society of London Series A, 370, 3217

\bibitem[{{Peter} \& {Bingert}(2012)}]{Peter2012}
{Peter}, H., \& {Bingert}, S. 2012, \aap, 548, A1

\bibitem[{{Pinto} {et~al.}(2016){Pinto}, {Gordovskyy}, {Browning}, \&
  {Vilmer}}]{Pinto2016}
{Pinto}, R.~F., {Gordovskyy}, M., {Browning}, P.~K., \& {Vilmer}, N. 2016,
  \aap, 585, A159

\bibitem[{{Snow} {et~al.}(2015){Snow}, {Botha}, \& {R{\'e}gnier}}]{Snow2015}
{Snow}, B., {Botha}, G.~J.~J., \& {R{\'e}gnier}, S. 2015, \aap, 580, A107

\bibitem[{{Snow} {et~al.}(2017){Snow}, {Botha}, {R{\'e}gnier}, {Morton},
  {Verwichte}, \& {Young}}]{Snow2017}
{Snow}, B., {Botha}, G.~J.~J., {R{\'e}gnier}, S., {et~al.} 2017, \apj, 842, 16

\bibitem[{{Srivastava} {et~al.}(2013){Srivastava}, {Botha}, {Arber}, \&
  {Kayshap}}]{Srivastava2013}
{Srivastava}, A.~K., {Botha}, G.~J.~J., {Arber}, T.~D., \& {Kayshap}, P. 2013,
  Advances in Space Research, 52, 15

\bibitem[{{Srivastava} {et~al.}(2010){Srivastava}, {Zaqarashvili}, {Kumar}, \&
  {Khodachenko}}]{Srivastava2010}
{Srivastava}, A.~K., {Zaqarashvili}, T.~V., {Kumar}, P., \& {Khodachenko},
  M.~L. 2010, \apj, 715, 292

\bibitem[{{Takasao} {et~al.}(2015){Takasao}, {Fan}, {Cheung}, \&
  {Shibata}}]{Takasao2015}
{Takasao}, S., {Fan}, Y., {Cheung}, M.~C.~M., \& {Shibata}, K. 2015, \apj, 813,
  112

\bibitem[{{Verwichte} {et~al.}(2009){Verwichte}, {Aschwanden}, {Van
  Doorsselaere}, {Foullon}, \& {Nakariakov}}]{Verwichte2009}
{Verwichte}, E., {Aschwanden}, M.~J., {Van Doorsselaere}, T., {Foullon}, C., \&
  {Nakariakov}, V.~M. 2009, \apj, 698, 397

\bibitem[{{Yuan} \& {Van Doorsselaere}(2016)}]{Yuan2016}
{Yuan}, D., \& {Van Doorsselaere}, T. 2016, \apjs, 223, 24

\end{thebibliography}

\listofchanges
   
\end{document}